\documentclass[preprint,aps,prd,superscriptaddress,nofootinbib]{revtex4-1}%

\usepackage{amsmath,amsfonts,amssymb,bm,graphicx,color,euscript}
\usepackage{dcolumn}% Align table columns on decimal point
\usepackage{bm}% bold math
\usepackage{hyperref}% add hypertext capabilities
\usepackage{epsfig}
\usepackage{mathrsfs}
\usepackage{slashed}
\def\sl{\!\!\!/}

\def\e{\epsilon}
\def\ve{\varepsilon}

\def\bs{\boldsymbol}
\def\nn{\nonumber}

\begin{document}
\preprint{}

%%%%%%%%%%%%%%%%%%%%%%%%%%%%%%%%%%%%%%%%%%%%
\title{Renormalization of the radiative jet function}
% Force line breaks with \\
%%%%%%%%%%%%%%%%%%%%%%%%%%%%%%%%%%%%%%%%%%%%
\author{Geoffrey~T.~Bodwin}
\email[]{gtb@anl.gov}
\affiliation{High Energy Physics Division, Argonne National Laboratory,
Argonne, Illinois 60439, USA}
\author{June-Haak~Ee}
\email[]{june\_haak\_ee@fudan.edu.cn}
\affiliation{Department of Physics, Korea University, Seoul 02841, Korea}
\affiliation{Key Laboratory of Nuclear Physics and Ion-beam Application (MOE) and Institute of Modern Physics, Fudan University, Shanghai 200433, China}
\author{Jungil~Lee}
\email[]{jungil@korea.ac.kr}
\affiliation{Department of Physics, Korea University, Seoul 02841, Korea}
\author{Xiang-Peng~Wang}
\email[]{xiangpeng.wang@anl.gov}
\affiliation{High Energy Physics Division, Argonne National Laboratory,
Argonne, Illinois 60439, USA}

%%%%%%%%%%%%%%%%%%%%%%%%%%%%%%%%%%%%%%%%%%%%
\date{\today}% It is always \today, today,
\begin{abstract}
We show how to compute directly the renormalization/evolution of the
radiative jet function that appears in the factorization theorems for
$B\to \gamma\ell\nu$ and $H\to \gamma\gamma$ through a $b$-quark
loop. We point out that, in order to avoid double counting of soft
    contributions, one should use in the factorization theorems a
    subtracted radiative jet function, from which soft contributions
    have been removed. The soft-contribution subtractions are zero-bin
subtractions in the terminology of soft-collinear effective theory.  We
show that they can be factored from the radiative jet function and that
the resulting soft-subtraction function gives rise to a nonlocal
renormalization of the subtracted radiative jet function. This is a
novel instance in which zero-bin subtractions lead to a nonlocality in
the renormalization of a subtracted quantity that is not present in
the renormalization of the unsubtracted quantity. We demonstrate the
use of our formalism by computing the order-$\alpha_s$ evolution kernel
for the subtracted radiative jet function. Our result is in agreement
with the result that had been inferred previously by making use of the
factorization theorem for $B\to \gamma\ell\nu$, but that had been
      ascribed to the unsubtracted radiative jet function.
\end{abstract}
\maketitle
%\tableofcontents
%%%%%%%%%%%%%%%%%%%%%%%%%%%%%%%%%%%%%%%%%%%%

\section{Introduction}

In the amplitudes for exclusive processes, contributions in which a
quark carries a soft momentum appear at subleading power in the ratio of
the quark mass to the large momentum transfer in the process.  These
contributions arise because there is a pinch singularity in the region
of soft quark momentum that has a subleading power dependence
\cite{Bodwin:2014dqa}. They are associated with endpoint singularities
in light-cone amplitudes.
 \cite{Beneke:2000ry,Beneke:2001at,Beneke:2001ev,Beneke:2003pa,
  Beneke:2003zv,Jia:2010fw,Benzke:2010js}.\footnote{There are also
analyses of corrections to {\it inclusive} cross sections at subleading
power in the inverse of the large momentum transfer. See, for example
Refs.~\cite{Beneke:2019oqx,Moult:2019mog,Moult:2019uhz,vanBeekveld:2019prq}.}
In the language of soft-collinear effective theory (SCET)
\cite{Bauer:2000yr,Bauer:2001yt,Beneke:2002ni,Bauer:2002nz,Beneke:2002ph},
the soft-quark contributions occur through a process in which a jet
function containing collinear quarks and gluons emits a soft quark via a
subleading-power interaction.  Such a jet function is called a radiative
jet function \cite{DelDuca:1990gz,Bonocore:2015esa,Bonocore:2016awd}.

A particular radiative jet function, which is the focus of this paper,
enters into the factorization theorem for the exclusive $B$-meson decay
$B\to \gamma\ell\nu$ \cite{Bosch:2003fc} and the factorization theorem
for the exclusive decay of the Higgs boson $H\to \gamma\gamma$ through a
$b$-quark loop \cite{Liu:2019oav}. 
\footnote{A discussion of the factorization theorem for the decay $B\to \gamma\ell\nu$ in the context of the method of regions is given in Ref. \cite{Wang:2016qii}.  Subleading-power corrections to the decay $B\to\gamma\ell\nu$ are discussed in Ref. \cite{Wang:2018wfj}.} In the remainder of this paper, we
will refer to this jet function as {\it the radiative jet function}.
The renormalization properties of the radiative jet
function are an essential ingredient in using these factorization
theorems to resum large logarithms of the ratios of $m_b/\mu$ or
$m_H/m_b$, where $\mu$ is the factorization scale, $m_b$ is the
$b$-quark mass, and $m_H$ is the Higgs-boson mass.

The radiative jet function has been computed through order $\alpha_s$ in
Ref.~\cite{Liu:2019oav}, and we have verified this calculation.  It
has also been computed through order $\alpha_s^2$ in
Ref.~\cite{Liu:2020ydl}.

The renormalization-group evolution of the radiative jet function in
order $\alpha_s$ has been inferred from the factorization theorem for
$B\to \gamma\ell\nu$, the renormalization-group invariance of the
physical amplitude for $B\to\gamma\ell\nu$, and the known
renormalization-group evolution kernel of the $B$-meson light-front
distribution, which also appears in the factorization theorem
\cite{Bosch:2003fc}.\footnote{This is an application of what is called
{\it the consistency condition} for the renormalization-group evolution
\cite{Bosch:2003fc}.}  That analysis has been extended to order
$\alpha_s^2$ in Ref.~\cite{Liu:2020ydl}. The renormalization of the
radiative jet function that is obtained from these analyses
is nonlocal in momentum space in that it involves
the convolution of a renormalization factor $Z_J$ with the
unrenormalized radiative jet function, rather than a simple multiplication.

\newpage

Although the renormalization properties of the radiative jet function
have been known indirectly for almost two decades, a method for
computing the nonlocal renormalization factor $Z_J$ directly from the
definition of the radiative jet function has remained elusive. In the words of
Ref.~\cite{Liu:2020ydl}, ``It is an embarrassment that there is no known
method in SCET to derive the anomalous dimensions of the jet functions
directly from their operator definitions.''

In this paper, we present a method to derive the anomalous dimension of
the radiative jet function directly from its operator definition. We
point out that the radiative jet function contains soft contributions
that are already taken into account in the soft functions of the
exclusive factorization theorems. These soft contributions must be
subtracted from the radiative jet function in order to avoid double
counting. Methods for the systematic subtraction of double-counted soft
contributions are familiar from the diagrammatic approach to
    factorization \cite{Collins:1981uk,Collins:2011zzd} and are known in
    SCET under the name {\it zero-bin subtractions}
    \cite{Manohar:2006nz}. We call the radiative jet function with the
soft contributions subtracted {\it the subtracted radiative jet
  function}. It is the subtracted radiative jet function, rather than
the radiative jet function, that should properly appear in the exclusive
factorization theorems.

We show that the soft subtractions can be factored from the radiative
jet function into a soft-subtraction function by making use of the
Grammer-Yennie approximation \cite{Grammer:1973db} and the graphical
Ward identities that are standard in diagrammatic factorization
\cite{Collins:1988ig}. The soft-subtraction function gives rise to
nonlocal ultraviolet (UV) divergences and accounts for all of the
nonlocal contributions in $Z_J$.

In dimensional regularization, soft subtractions (zero-bin subtractions)
generally result in scaleless integrals that can be interpreted as being
proportional to a difference between UV and infrared (IR) poles. The
soft subtractions then have the function of converting IR poles to UV
poles. In a fixed-order calculation, if one does not distinguish IR
poles from UV poles, then the soft subtractions do not affect the
result. As we will see, this is also the case for the soft subtractions
of the radiative jet function. However, beyond one-loop order, the UV
poles of the soft subtractions are proportional to nonlocal convolutions
over light-front momenta, and, so, they lead to nonlocal contributions
to $Z_J$. To the best of our knowledge, this is the first time that a
nonlocal renormalization owing to the effect of zero-bin subtractions
has been observed.

The remainder of this paper is organized as follows.  In
Sec.~\ref{sec:notation-defs}, we present some of the notation and
conventions that we use throughout this paper. In
Sec.~\ref{sec:radiative-jet-function}, we give the operator definition of the
radiative jet function and show how to factor the radiative jet function
into a convolution of a soft-subtraction function and a subtracted
radiative jet function.  The renormalization procedure for the
subtracted radiative jet function is outlined in
Sec.~\ref{sec:renorm}. In Sec.~\ref{sec:LO-exp}, we record the
leading-order expressions for the radiative jet function, the
soft-subtraction function, the subtracted radiative jet function,
    and the soft-subtraction renormalization. In
Sec.~\ref{sec:nlo}, we present the application of our method to the
renormalization of the subtracted radiative jet function in order
$\alpha_{s}$. Our result for the order-$\alpha_s$ renormalization-group
kernel of the subtracted radiative jet function is in agreement with
the result in Ref.~\cite{Bosch:2003fc}, although, in that work, the
    renormalization-group kernel is ascribed to the unsubtracted
    radiative jet function.  In Sec.~\ref{sec:softvanish}, we argue
that soft subtractions generally vanish in dimensional regularization if
one does not distinguish between UV and IR divergences, and, hence, do
not affect existing fixed-order calculations of the radiative jet
function. Finally, we summarize and discuss our results in
Sec.~\ref{sec:summary}.

\begin{figure}
\begin{center}
\includegraphics[width=0.25\columnwidth]{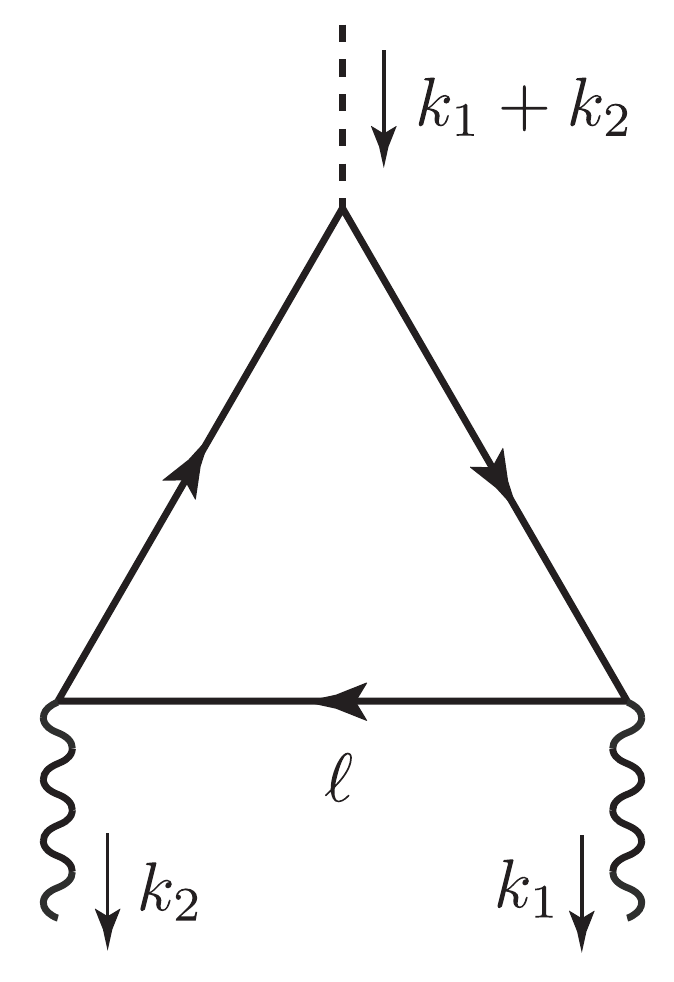}
\caption{$H\rightarrow b\bar{b}\rightarrow \gamma\gamma$ at leading order. 
The dashed line represents the Higgs boson, the solid line represents
    the $b$ quark, and the wavy lines are the photons.
\label{fig:H2gamma}%
}
\end{center}
\end{figure}

\section{Notation and conventions\label{sec:notation-defs}}

In this section, we establish some of our notation and conventions.

The lowest-order contribution to $H\to \gamma\gamma$ through a $b$-quark
loop is shown in Fig.~\ref{fig:H2gamma}, 
which establishes our conventions for the
momenta of the external particles and the orientation of the
internal quark loop.

We adopt dimensional regularization in $D=4-2\epsilon$ space-time
    dimensions to regularize the divergences in the loop integrations.

We define two light-like vectors, $n$ and $\bar{n}$, which satisfy the
conditions
%---------------
\begin{eqnarray}
%---------------
n^2=0, \qquad \bar{n}^2=0, \qquad n\cdot \bar{n}=2.
%---------------
\end{eqnarray}
%---------------
Any four-vector $\ell$ can be decomposed, in terms of these light-like
vectors, as
%---------------
\begin{eqnarray}
%---------------
\ell^{\mu} 
= 
(n\cdot \ell)\frac{\bar{n}^{\mu}}{2} 
+ (\bar{n}\cdot \ell)\frac{n^{\mu}}{2} 
+ \ell_{\perp}^{\mu} 
= 
\ell_{+}\frac{\bar{n}^{\mu}}{2} 
+ \ell_{-}\frac{n^{\mu}}{2} 
+ \ell_{\perp}^{\mu}.
%---------------
\end{eqnarray}
%---------------
Here, $\ell_\perp=(0,\ell^1,\ell^2,\cdots,\ell^{D-2},0)$ is a
$(D-2)$-dimensional vector, and we have defined
%---------------
\begin{eqnarray}
%---------------
\ell_{+}\equiv n\cdot \ell,  
\quad
\ell_{-} \equiv \bar{n}\cdot \ell.
%---------------
\end{eqnarray}
%---------------
%It is customary to represent a momentum in these light-cone coordinates by 
%\begin{eqnarray}
%\ell^{\mu} = (\ell_{+},\ell_{-},\ell_{\perp}).
%\end{eqnarray}
As a consequence, the scalar product of two vectors $k$ and $\ell$ becomes
%---------------
\begin{eqnarray}
%---------------
k\cdot \ell 
&=& 
\frac{k_{+}\ell_{-}}{2}+\frac{k_{-}\ell_{+}}{2} +k_{\perp}\cdot \ell_{\perp} 
= 
\frac{k_{+}\ell_{-}}{2}+\frac{k_{-}\ell_{+}}{2}
-\bs{k}_{\perp}\cdot \bs{\ell}_{\perp},
%---------------
\end{eqnarray}
%---------------
where $\bs{\ell}_{\perp}$ and $\bs{k}_{\perp}$ are $(D-2)$-dimensional
Euclidean vectors.  

We make use of the following SCET notations.  $q_s$ is the soft-quark
Dirac spinor, $G_s=T^a G_s^a$, where $G_s^a$ is the soft-gluon field
with color index $a$ and $T^a$ is an $SU(3)$ color matrix in the
fundamental representation, $G_n=T^a G_n^a$, where $G_n^a$ is an
$n$-hard-collinear gluon field with color index $a$, $A_n$ is an
$n$-hard-collinear photon field, $\xi_n$ is an $n$-hard-collinear-quark
Dirac spinor, which satisfies $\slashed{n}\xi_n=0$, $iD_{n}^{\perp\mu} =
i\partial^{\perp\mu}+g_{s}G_{n}^{\perp\mu}+e_{q}A_{n}^{\perp\mu}$ is a
transverse covariant derivative, $g_s=\sqrt{4\pi\alpha_s}$ is the strong
coupling, and $e_q$ is the electric charge of the collinear quark.
    
The scaling of a soft momentum is given by
\begin{equation}
\label{eq:soft-scaling}
k_{s}^{\mu}\sim (\lambda,\lambda,\lambda)Q.
\end{equation}
Here $Q$ is the large momentum scale and $\lambda\sim m_b/Q$. 
We follow the convention
\begin{equation}
k=(k_+,k_-,\bm{k}_\perp)
\end{equation}
for the light-front coordinates.
The scaling of an $n$-hard-collinear momentum is given by
\begin{equation}
\label{eq:hard-collinear-scaling}
k_{n}^{\mu}\sim (\lambda,1,\lambda^{\frac{1}{2}})Q.
\end{equation}
The soft-quark spinor scales as $\lambda^{\frac{3}{2}}$, the
soft-gluon field scales as $\lambda$, the $n$-hard-collinear spinors
scale as $\lambda^{\frac{1}{2}}$, and the $n$-hard-collinear gauge
fields scale as the $n$-hard-collinear momentum $k^\mu_n$ in
Eq.~(\ref{eq:hard-collinear-scaling}) \cite{Beneke:2002ph}.

\section{Radiative jet function
\label{sec:radiative-jet-function}}
\subsection{Operator definition of the radiative jet function }

In the factorization theorem of Ref.~\cite{Liu:2019oav}, the
interactions of virtual hard-collinear quarks and gauge bosons with an
outgoing real hard-collinear photon are described by the
radiative jet function. The radiative jet function couples to the
soft quark by virtue of an interaction between the soft quark and the
hard-collinear quarks and gauge fields that first appears at subleading
order in the expansion of the SCET Lagrangian.  This coupling is
given, in the notation of \cite{Beneke:2002ph}, by
\begin{equation}
\label{subleadingL}
   {\cal L}_{q\,\xi_n}^{(1/2)}(x) = \bar
   q_s(x_-)\,W_n^{\dagger}(x)\,i\slashed{D}_n^\perp\,\xi_n(x) +
   \mbox{H.c.},
\end{equation}
where H.c. denotes the Hermitian-conjugate contributions.  Here,
the argument of the soft-quark field $q_s$ is taken to be $x_-$, with
$x_-^\mu = (\bar{n}\cdot x)\frac{n^\mu}{2}$. This argument effects the
expansion in momentum space at leading power in $\lambda$, in which the
collinear subdiagrams depend only on the $+$ component of the soft
momentum. (This is {\it the multipole expansion} of SCET
\cite{Beneke:2002ph}.)  $W_{n}$ is a collinear Wilson line, which is
defined by
\begin{equation} 
\label{eq:collinear-wilson}
W_{n}(x) = P\,
   \textrm{exp}\left[ig_{s}\int_{-\infty}^{0}ds\,\bar{n}\cdot
     G_{n}(x+s\bar{n})+ie_{q}\int_{-\infty}^{0}ds\,\bar{n}\cdot
     A_{n}(x+s\bar{n})\right], 
\end{equation}
where $P$ denotes the path ordering of the exponential.
We drop the term involving $A_n$ in the remainder of this paper.

In discussing the soft subtraction, we will 
encounter the soft Wilson line along $n$ which is defined by
%---------------
\begin{eqnarray}
%---------------
\label{def:Wilson-line}%
S_{n}(x)&=& P\,\textrm{exp}\left[ig_{s}\int_{-\infty}^{0}dt\, n\cdot G_{s}(x+tn)\right].
%---------------
\end{eqnarray}
%---------------

We take the definition of the radiative jet function $J(p^{2})$ that is
given in Eq.~(1.2) of Ref.~\cite{Liu:2020ydl}:
\begin{eqnarray}
\label{defJF}%
\bar{J}(\bar n\cdot p,p^2)&\equiv&
i\int d^{D}x\,e^{i\frac{\ell_+ x_-}{2}}\langle \gamma(k_1)  | T\,
\big(W^{\dagger}_{n}i\slashed{D}^{\perp}_{n}\xi_{n}\big)^{a}(x)\big(\bar{\xi}_{n}W_{n}\big)^{b}(0)
| 0\rangle \nn\\
&=&
ie_{q}\delta^{ab}\,\rlap/\slashed{\varepsilon}^{*}_{\perp}(k_1)\frac{\slashed{n}}{2}\frac{i\bar{n}\cdot
  p }{p^{2} +{i\varepsilon}}J(p^{2}).
\end{eqnarray}
Here, we have made the color indices $a$ and $b$ explicit, and we have
defined $\bar{J}(\bar n\cdot p,p^2)$ to be the complete expression in
Eq.~(\ref{defJF}).  $\gamma(k_1)$ denotes a real photon with momentum
$k_1=(0,k_{1-},\bm{0}_\perp)$ and polarization $\varepsilon^*(k_1)$, and
$T$ denotes the time-ordered product.  We have inserted a factor $i$ in
front of the matrix element in Eq.~(\ref{defJF}), so that there is a
factor $i$ that is associated with each real-photon vertex, whether it
arises from the covariant derivative in the matrix element or from the
QED interaction Lagrangian. $p$ is defined by
\begin{equation}
p^{\mu}\equiv k_1^{\mu} +\ell^{\mu},
\end{equation}
and $\ell$ has the interpretation of the soft momentum
that is carried by the soft quark in Eq.~(\ref{subleadingL}).
Note that
\begin{equation}
\ell_+=p_+.
\end{equation}
Throughout this paper, we take the approximation 
\begin{equation}
\ell^\mu \approx\ell_{+}\frac{\bar{n}^{\mu}}{2}.
\label{eq:ell-approx}
\end{equation} 
Since $k_1$ satisfies hard-collinear scaling and $\ell$ satisfies soft
scaling, this approximation gives the leading power in $\lambda$ in the
argument $p^2$ of the radiative jet function. This is the multipole
expansion to which we alluded earlier. 
It follows that 
\begin{equation}
p^2\approx k_{1-} \ell_+= p_-p_+.
\label{eq:p-sq-approx}%
\end{equation}
This approximation has been invoked in writing $p^2$ in the arguments of
$\bar J$ and $J$.  In the remainder of this paper, we suppress the
argument $\bar n\cdot p \approx k_{1-}$ in $\bar J$.

$\bar{J}(p^2)$ contains contributions in which the real photon attaches
to the covariant derivative and contributions in which the real photon
attaches to the quark line. Following Ref.~\cite{Liu:2019oav}, we call
the former contributions $\bar{J}_A(p^2)$, and we call the latter
contributions $\bar{J}_G(p^2)$, writing
\begin{equation}
\label{eq:JA-JG}
\bar{J}(p^2)=\bar{J}_A(p^2)+\bar{J}_G(p^2). 
\end{equation}

In computing contributions to the radiative jet function in this
    paper, we do not use the SCET Feynman rules. Instead, we follow a
    procedure that is equivalent, but more amenable to a graphical
    analysis. Starting from the Feynman rules for QCD, we insert
projectors 
\begin{eqnarray}
P_n&=&\frac{\slashed{n}\slashed{\bar {n}}}{4},
\nn\\
~
P_{\bar{n}} &=&\frac{\slashed{\bar
    {n}}\slashed{n}}{4} 
\end{eqnarray}
on the outgoing and incoming ends of the quark lines, respectively, so
    as to obtain the components of the Dirac spinor that are large in
    $n$-hard-collinear scaling and that correspond to the
    $n$-hard-collinear spinors $\xi_n$ and $\bar \xi_n$, respectively.

\subsection{Subtraction and factorization of soft contributions to the
  radiative jet function}
  \label{subsec:subtraction-factorization}

The radiative jet function, as defined in Eq.~(\ref{defJF}), also
contains soft contributions. These must be removed in order to avoid
double counting of contributions in the soft function in the
factorization theorems. One can factor the soft contributions from the
jet function by making use of standard techniques from the diagrammatic
methods for proving factorization theorems \cite{Collins:1989gx}.  We
carry out the factorization in the Feynman gauge. However, the resulting
expressions are gauge invariant. Our approach in dealing with the soft
    subtractions is analogous to the one that is given in Sec.~10.8.7 of
    Ref.~\cite{Collins:2011zzd}.

Soft divergences can develop if one end of a soft gluon attaches either
to a Wilson line or to a soft quark line and the other end attaches to
an $n$-hard-collinear line. There is no soft divergence if both ends of
a gluon attach to $n$-hard-collinear lines. If both ends of a soft
gluon attach to a soft quark line, then that soft divergence is part of
the soft function or is internal to a soft subtraction, and no special
treatment of it is required.

Let us consider first the case of soft gluons that attach to
hard-collinear lines, but not to the Wilson lines $W_n$ and
$W^\dagger_n$. At leading order in the scaling parameter $\lambda$, the
current $j^\mu$ in the hard-collinear lines to which the soft gluons
couple is proportional $n^\mu$.  It follows that, in the attachment of
that soft gluon to a hard-collinear line, its polarization sum
$g^{\mu\nu}$ can be replaced with $k^\nu n^\mu/(n\cdot k+i\ve)$, where
the $i\ve$ prescription corresponds to a momentum routing in which $k$
flows parallel to the arrow on the quark propagator. Note that the
    polarization now corresponds to a pure gauge.  This is the
Grammer-Yennie approximation \cite{Grammer:1973db}. Then, one can apply
graphical Ward identities to show that sum over the attachments of all
such soft gluons to the hard-collinear lines can be replaced with the
sum over all attachments of the soft gluons to Wilson lines $S_n$ and
$S^\dagger_n$ that attach to the quark line immediately to the
outgoing side and immediately to the incoming side,
respectively, of the outermost hard-collinear-gluon attachments. Details
of this step are given following Eq.~(4.3) of
Ref.~\cite{Collins:1988ig}. In SCET, this step can be implemented by
making use of field redefinitions \cite{Bauer:2000yr}.

In carrying out this analysis, we omit diagrams that contain
$n$-hard-collinear subdiagrams that are not connected by lines carrying
$n$-hard-collinear momenta to the external photon.  These diagrams lead
to contributions that are not properly part of the radiative jet
function because the disconnected subdiagram does not yield a pinch
singularity in the $n$-hard-collinear momentum region.  (At one-loop
order, these contributions vanish in dimensional regularization.)  Such
diagrams contain quark lines that carry soft momenta and, so, if they
were to contribute to the radiative jet function, they would cause the
Grammer-Yennie approximation and the factorization of the soft
subtractions to fail.  We will discuss an example of such a diagram in
Sec.~\ref{sec:J-G-alphas}.
  
Next, let us consider the case of soft gluons that attach to the
collinear Wilson lines $W_n$ and $W_n^\dagger$. Here, we use the fact
that soft-gluon attachments to the collinear Wilson lines $W_n$ and
$W_n^\dagger$ that lie to the outside of all collinear-gluon attachments
factor into new soft Wilson lines $S_{\bar n}$ and $S_{\bar n}^\dagger$,
respectively, that lie to the outside of the $W_n$ and $W_n^\dagger$
collinear Wilson lines of the jet function
\cite{Collins:1988ig}.\footnote{Note that the collinear Wilson line
$W_n$ contains projections of the gluon field onto $\bar{n}$.  [See
  Eq.~(\ref{eq:collinear-wilson}).] Hence, $W_n$ gives rise to $S_{\bar
  n}$, and $W_n^\dagger$ gives rise to $S_{\bar n}^\dagger$.}
Soft-gluon attachments to the collinear Wilson lines $W_n$ and
$W_n^\dagger$ that lie to the inside of collinear-gluon attachments
yield power-suppressed contributions \cite{Collins:1988ig}.

\begin{figure}
     \includegraphics[width=1.0\columnwidth]{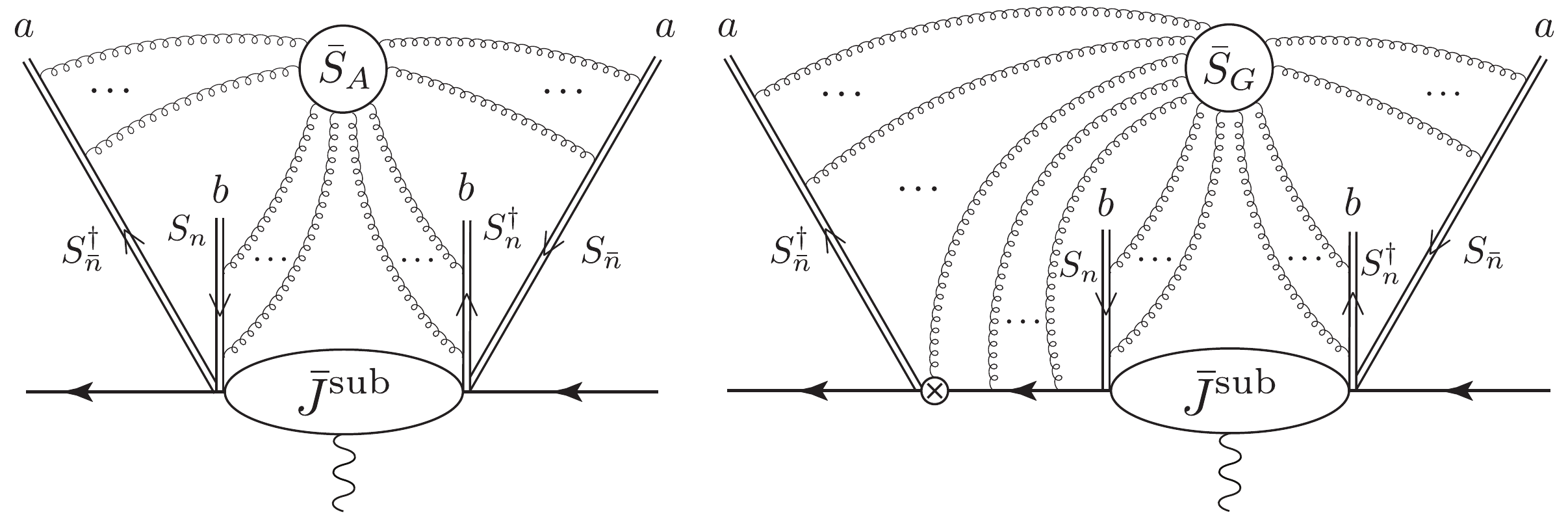}
    \caption{Diagrammatic representation of the extraction of the soft
          contributions from $\bar J$. $a$ and $b$ are color
          indices.}
     \label{fig:jetsub}
\end{figure}

The resulting diagrams have the form of those in
    Fig.~\ref{fig:jetsub}. 
At this stage, the blobs labeled $\bar J^\textrm{sub}$ contain all of
the collinear contributions and contain no soft contributions. The first
diagram in Fig.~\ref{fig:jetsub}, whose soft subtractions we denote by
$\bar S_A$, arises from the contributions in which no soft gluons attach
to the quark line to the outgoing side of the collinear-gluon
attachments. The second diagram in Fig.~\ref{fig:jetsub}, whose soft
subtractions we denote by $\bar S_G$, arises from the contributions in
which one or more soft gluons attach to the quark line to the outgoing
side of the collinear-gluon attachments.\footnote{Note that there is
    not a one-to-one correspondence between $\bar S_A$ and $\bar J_A$ or
    between $\bar S_G$ and $\bar J_G$.  $\bar J_A$ decomposes into $\bar
    S_A\otimes \bar J_A^{\rm sub}$, but $\bar J_G$ decomposes into $\bar
    S_A\otimes \bar J_G^{\rm sub}$, $\bar S_G\otimes \bar J_G^{\rm
      sub}$, and $\bar S_G\otimes \bar J_A^{\rm sub}$.}

We call the quark line that lies between the covariant derivative and
$S_n$ in the second diagram in Fig.~\ref{fig:jetsub} {\it the
      $n$-hard-collinear-soft quark line}. Its Feynman rules are
obtained from those for the $n$-hard-collinear quark line by retaining
only those contributions that are leading in the soft scaling of the
gluon momenta.  An $n$-hard-collinear quark line couples to a soft quark
line only through the covariant derivative or $W_n^\dagger$ in
Eq.~(\ref{subleadingL}), and at least one $n$-hard-collinear gluon must
attach to the covariant derivative or to $W^\dagger_n$ in order to
produce an $n$-hard-collinear momentum in the quark line. Therefore, an
$n$-hard-collinear-soft quark line couples to a soft quark line only
through the covariant derivative or $S^\dagger_{\bar n}$, and at least
one soft gluon must attach to the covariant derivative or to
$S^\dagger_{\bar n}$.

The blob in the first diagram of Fig.~\ref{fig:jetsub} is in the
form of a radiative jet function, except that it also contains soft
    subtractions, which can be implemented by subtracting the
    Grammer-Yennie form from specific gluon vertices.  The blob in the
second diagram of Fig.~\ref{fig:jetsub} requires some re-arrangement to
put it into the form of a radiative jet function (with soft
subtractions). One approach is simply to invoke the form of the SCET
Lagrangian for $n$-hard-collinear quarks and gluons and their couplings
to a soft quark and a real photon to deduce that the blob in the second
diagram of Fig.~\ref{fig:jetsub} takes the form of a radiative jet
    function (with soft subtractions) in SCET.  Here, it is important
that we have defined $\bar J$ in Eq.~(\ref{defJF}) in such a way that
the vertex for a transverse photon is always $ie_{q}\gamma_\perp^\mu$,
regardless of whether the photon attaches to a quark line or to a
covariant derivative. In Appendix~\ref{app:blob-rearrangement}, we
sketch how the re-arrangement of the blob in the second diagram of
    Fig.~\ref{fig:jetsub} into the radiative-jet form can be achieved
in the diagrammatic approach.

We remind the reader that, owing to the multipole expansion, we keep
only the plus components of soft momenta that enter $\bar J^{\rm
  sub}$. Using this fact, we can express the diagrams of
Fig.~\ref{fig:jetsub}, as convolutions over the plus component of
momentum:
\begin{equation}
\label{eq:modj}%
\bar J(p_-p_+)=\int \frac{dk_+}{p_+} \bar S(k_+) 
\bar J^{\rm sub}\left[p_-(p_+-k_+)\right]
\equiv \bar S \otimes \bar J^{\rm sub}.
\end{equation}
Note that we can also write this convolution in the form
\begin{equation}
\bar S \otimes \bar J^{\rm sub}= \int dx\, \bar S[(1-x)p_+] 
\bar J^{\rm sub}(xp^2),
\end{equation}
where we have made the variable change
\begin{equation}
\label{eq:kp-in-terms-of-x}
k_+=(1-x)p_+,
\end{equation}
and we have used $p^2\approx p_+p_-$
[Eq.~(\ref{eq:p-sq-approx})]. Because both $\bar S$ and $\bar J^{\rm
  sub}$ are Dirac matrices, the order of the factors in the convolution
is significant.

The soft-subtraction function is given by
\begin{subequations}%
\label{eq:softsub}
\begin{equation}
\label{eq:modj-2}%
\bar{S}(k_+)=\bar S_A(k_+)+\bar S_G(k_+),
\end{equation}
where
\begin{eqnarray}
\bar{S}_A(k_+)&=&
\frac{p_+}{2\pi N_c}
\int\frac{dx_-}{2}\,e^{i\frac{k_+ x_-}{2}} 
\langle 0|T(S_{\bar n}^{\dagger} S_n)^{ab}(x_-) 
(S_n^{\dagger} S_{\bar n})^{ba}(0)|0\rangle,\nn\\
\bar{S}_G(k_+) &=&
\frac{ip_+}{2\pi N_c}
\int d^Dy\, 
\int \frac{dx_-}{2}\,
e^{i\frac{p_+y_-}{2}} 
e^{i\frac{(k_+-p_+)x_-}{2}}  
\nonumber \\
&&
\times
\langle 0|T(S_{\bar n}^{\dagger}P_n i\slashed{D}_{s}^\perp
\psi_{n,s})^a(y) (\bar \psi_{n,s}P_nS_n)^b(x_-)
(S_n^{\dagger} S_{\bar n})^{ba}(0)|0\rangle.
\end{eqnarray}
\end{subequations}%
In the arguments of $\bar{S}_A$ and $\bar S_G$, we have suppressed the
dependences on the parameter $p_+=\ell_+$. The covariant derivative is
labeled with a subscript $s$ as a reminder that the gauge fields and
their momenta in the covariant derivative satisfy soft scaling.
$\psi_{n,s}$ is the $n$-hard-collinear-soft field, which corresponds to
the $n$-hard-collinear-soft line in the second diagram in
Fig.~\ref{fig:jetsub} that we have described earlier. The subscript
$n,s$ is a reminder that the propagator and interactions of this field
with soft gauge fields are obtained by first taking an approximation in
which the quark momentum and the momenta of the attached gauge fields
have $n$-hard-collinear scaling and then taking an approximation in
which the quark momentum and the momenta of the attached gauge fields
have soft scaling. One consequence of this is that the propagator for
$\psi_{n,s}$ is massless.

In the equation for $\bar S_G$, the left-hand factor $P_n$ arises from
the factor $P_n$ in the unsubtracted radiative jet function $\bar J$,
while the right-hand factor $P_n$ arises from applying the identity
$P_n=P_n^2$ to the factor $P_n$ on the left side of $\bar J^{\rm sub}$
in Eq.~(\ref{eq:modj}).  As we have mentioned, at least one soft gluon
must attach to the covariant derivative or to $S_{\bar n}^\dagger$ in
$\bar S_G$. Consequently, one should subtract the non-interacting
part, that is, make the replacement $S_{\bar n}^\dagger i \slashed
D_{s}^\perp\to S_{\bar n}^\dagger i \slashed D_{s}^\perp-i\slashed
\partial^\perp$.

At any order in perturbation theory, the soft-subtraction contributions
can also be obtained by starting with the jet contributions and
retaining only the parts that have leading soft scaling
[Eq.~(\ref{eq:soft-scaling})]. However, in this form, it is not apparent
that the soft subtractions factor from the radiative jet function to
yield the form in Eq.~(\ref{eq:softsub}). As we will see, it is the form
in Eq.~(\ref{eq:softsub}) that leads to the nonlocality in the
renormalization of the radiative jet function.

It follows from Eq.~(\ref{eq:modj}) that $\bar{J}^{\rm sub}$ is
given by
\begin{eqnarray}
\label{eq:jsub}
\bar J^{\rm sub}&=&\bar S^{-1}\otimes \bar J.
\end{eqnarray}
The quantity $\bar S^{-1}$ can be obtained to any order in $\alpha_s$ by 
making use of the expansion of $\bar S$ in powers of $\alpha_s$, namely,
\begin{eqnarray}
\bar S&=&1+\alpha_s \bar S^{(1)}+\ldots 
\end{eqnarray}
and solving the equation  
\begin{eqnarray}
\bar S^{-1}\otimes \bar S&=&1
\end{eqnarray}
iteratively. Specifically, in order $\alpha_s$, Eq.~(\ref{eq:jsub}) can
be written in the form
\begin{eqnarray}
\bar J^{\rm sub}&=&(1-\alpha_s \bar S^{(1)})\otimes \bar J.
\end{eqnarray}

\section{Renormalization of the radiative jet function}
\label{sec:renorm}

The renormalization of $\bar J^{\rm sub}$ arises from two
  sources:  the radiative jet function and the soft function.

The radiative jet function $\bar{J}$ in Eq.~(\ref{eq:modj})
is multiplicatively renormalized:
\begin{equation}
\bar J_R(p^2;\mu)=Z_{\bar J}(p^2;\mu) \bar J(p^2),
\end{equation}
where $\mu$ is the renormalization scale.  That is, the renormalization
of $\bar J$ is local. This follows from the fact, in order for there to
be a nonlocal renormalization for $\bar{J}$, a UV-divergent diagrammatic
loop must transfer a plus component of loop momentum $k_+$ from one
external leg of $\bar J$ to the other external leg of $\bar J$.
However, a loop that transfers $k_+$ in this way does not have a
UV-divergent power count because there are too many propagators in the
loop. We will see explicit examples of this in the one-loop calculations
in Sec.~\ref{sec:nlo}. 

The soft-subtraction function $\bar S$ is nonlocally renormalized:
\begin{equation}
\label{eq:Z-bar-J-conv}%
\bar S(k_+)=
\int \frac{dk_+'}{k_+}
\bar S_R(k_+-k_+';\mu) Z_{\bar S}^{-1}(k_+'/k_+;\mu)
\equiv 
\bar S_R\otimes Z_{\bar S}^{-1}.
\end{equation}
In contrast with $\bar J$, $\bar S$ can have nonlocal renormalizations
    because two of the operators in the definitions of $\bar S$ in
    Eq.~(\ref{eq:softsub}) are separated only in the minus
    direction. That is, in momentum space, only the plus component of a
    loop momentum that routes through these operators is
    constrained. Consequently, a loop momentum $k$ can transfer $k_+$
    from one external leg of $\bar S$ to the other external leg, and the
    integration over $\bm{k}_\perp$ can still be UV divergent. We will
    also see explicit examples of this phenomenon in the one-loop
    calculation in Sec.~\ref{sec:nlo}.

Note that we can also write the convolution in the form
\begin{equation}
\label{eq:Z-bar-J-conv-2}%
\bar S(k_+)=
\int dx' \,\bar S_R (x'k_+;\mu) Z_{\bar S}^{-1}(1-x';\mu),
\end{equation}
where we have used $k'_+=(1-x')k_+$.  Under the change of integration
variables $x'=1-x$, the convolution in Eq.~(\ref{eq:Z-bar-J-conv}) is
identical to the one in Eq.~(\ref{eq:modj}), aside from a trivial
rescaling of the argument of $Z_{\bar S}^{-1}$ with a factor of $p^2$
and the argument of $\bar S_R$ with a factor $p_+/k_+$. Therefore, we
use the same notation ($\otimes$) for both convolutions. However, one
should keep in mind the rescalings of arguments that are implicit in
this notation.  It follows that
\begin{eqnarray}
\bar J^{\rm sub}&=&\bar S^{-1}\otimes \bar J
=(\bar S_R\otimes Z_{\bar S}^{-1})^{-1}\otimes Z_{\bar J}^{-1} \bar J_R
=Z_{\bar J}^{-1} Z_{\bar S}\otimes \bar S_R^{-1}\otimes \bar J_R\nn\\
&=&Z_{\bar J}^{-1} Z_{\bar S}\otimes \bar J_R^{\rm sub}.
\end{eqnarray}
This implies that
\begin{eqnarray}
\bar J_R^{\rm sub}(p_-p_+;\mu)&=&
(Z_{\bar J^{\rm sub}}\otimes \bar J^{\rm sub})(p_-p_+;\mu)\nn\\
&=&\int \frac{dk_+}{p_+}\, 
Z_{\bar J^{\rm sub}}(k_+/p_+,p^2;\mu) 
\bar J^{\rm sub}[p_-(p_+-k_+)],
\end{eqnarray}
or, equivalently,
\begin{subequations}
\label{eq:Jsub-renorm}%
\begin{equation}
\bar J_R^{\rm sub}(p^2;\mu)
=\int dx\, Z_{\bar J^{\rm sub}}[(1-x),p^2;\mu] \bar J^{\rm sub}(xp^2),
\end{equation}
where
\begin{equation}
Z_{\bar J^{\rm sub}}[(1-x),p^2;\mu]=Z_{\bar J}(p^2;\mu) 
Z_{\bar S}^{-1}[(1-x);\mu],
\label{eq:Jsub-renorm-Z}%
\end{equation}
\end{subequations}%
and we have used $p^2\approx p_+p_-$.  Note
that, the renormalization factors $Z_{\bar J}$ and $Z_{\bar S}^{-1}$
contain only those renormalizations that are associated with the
operator matrix element (operator renormalizations) and do not contain
the mass and coupling-constant renormalizations of QCD, except for the
quark wave-function renormalization.

In order to make contact with the renormalization of the radiative jet
    function in Ref.~\cite{Liu:2020ydl}, we make use of
    Eq.~(\ref{eq:Jsub-renorm}) and the relation between $J(p^2)$ and
    $\bar{J}(p^2)$ in Eq.~(\ref{defJF}) to obtain
\begin{subequations}%
\begin{equation}
J_R^{\rm sub}=Z_{J^{\rm sub}}\otimes J^{\text{sub}},
\end{equation}
where
\begin{equation}
\label{eq:Z-J-sub}%
Z_{J^{\rm sub}}[(1-x),p^2;\mu]=(1/x) Z_{\bar J^{\rm sub}}
[(1-x),p^2;\mu].
\end{equation}
\end{subequations}%

\section{Lowest-order expressions 
%for $\pmb{\bar{J}}^{\textbf{sub}}$($\pmb{p}^{\textbf{2}}$)
}
\label{sec:LO-exp}
From the definition of $\bar{J}$ in Eq.~(\ref{defJF}), it follows that
the lowest-order expression for $\bar{J}$ is given by
\begin{equation}
\bar{J}^{(0)}(p^2)
=
ie_{q}\delta^{ab}\,\rlap/\slashed{\varepsilon}^{*}_{\perp}(k_1)\frac{\slashed{n}}{2}\frac{i\bar{n}\cdot
  p}{p^{2} +{i\varepsilon}} J^{(0)}(p^2),
\end{equation}
where $J^{(0)}(p^{2})=1$ is the contribution to $J(p^2)$ at leading
order in $\alpha_s$.  From the definition of the soft-subtraction
    function in Eq.~(\ref{eq:softsub}), we find that the
lowest-order expressions for $\bar{S}_A$ and $\bar{S}_G$ are given by
\begin{eqnarray}
\label{eq:lowest-order-S}%
\bar{S}_A^{(0)}(k_+)&=&p_+\delta(k_+),
\nonumber \\
\bar{S}_G^{(0)}(k_+) &=&
0.
\end{eqnarray}
Then, making use of the relation in Eq.~(\ref{eq:modj}),
we find that the lowest-order expression for $\bar J^{\rm sub}$ 
is the same as  $\bar J^{(0)}$:
\begin{eqnarray}
\label{eq:j-bar-sub-0}%
\bar J^{(0)}(p^2)
&=&\int \frac{dk_+}{p_+}\bar{S}_A^{(0)}(k_+)
 \bar J^{\rm sub(0)}\left[p_-(p_+-k_+)\right]
=\bar J^{\rm sub(0)}(p^2).
\end{eqnarray}

Note also that 
\begin{eqnarray}
\label{eq:lowest-order-S-conv}%
[\bar S^{(0)}_R\otimes (Z_{\bar S}^{-1})^{(1)}]
\otimes \bar J^{\rm sub}
&=&(Z_{\bar S}^{-1})^{(1)}\otimes \bar J^{\rm sub}\nn 
=\int \frac{dk_+}{p_+} (Z_{\bar S}^{-1})^{(1)}(k_+/p_+) \bar J^{\rm
  sub}[p_-(p_+-k_+)]\nn\\
&=&\int dx(Z_{\bar S}^{-1})^{(1)}(1-x) \bar J^{\rm sub}(xp^2),
\end{eqnarray}
where we have used $\bar S^{(0)}_R=\bar{S}_A^{(0)}$ in the first
equality and used $k_+=(1-x)p_+$ and $p^2\approx p_+p_-$ 
in the last equality. We make use of
Eq.~(\ref{eq:lowest-order-S-conv}) in computing the order-$\alpha_s$
contribution to $Z_{\bar S}^{-1}$ in Sec.~\ref{sec:nlo}.

Finally, we have for the lowest-order contribution to $Z_{\bar S}^{-1}$  
\begin{equation}
\label{eq:LO-Z-bar-S}%
(Z_{\bar S}^{-1})^{(0)}(1-x)=\delta(1-x).
\end{equation}

%%%%%%%%%%%%%%%%%%%%%%%%%%%%%%%%%%%%%%%%%%%%%%%%%%%%%%%%%%%%%%%%%%%%%%%%%%%%%%%%%%%%%%%%%%%%%%%%%%
\section{Renormalization of $\pmb{\bar{J}}^{\textbf{sub}}$($\pmb{p}^{\textbf{2}}$) at order $\pmb{\alpha_s}$}
\label{sec:nlo}

In this section, we compute the renormalization of the subtracted
    radiative jet function in order $\alpha_S$.

$Z_{\bar S}$ and $Z_{\bar J}$ have expansions in powers of
$\alpha_s$:
\begin{subequations}%
\begin{eqnarray}
Z_{\bar S}&=&\bm{1}+\alpha_s Z_{\bar S}^{(1)}+\ldots , \\
Z_{\bar J}&=&1+\alpha_s Z_{\bar J}^{(1)}+\ldots.
\end{eqnarray}
\end{subequations}%
Here, $\bm{1}$ represents the lowest-order contribution to $Z_{\bar
      S}$ [Eq.~(\ref{eq:LO-Z-bar-S})].  Then, from
      Eq.~(\ref{eq:Jsub-renorm-Z}) we find that, to order $\alpha_s$
  in the renormalizations, $\bar J_R^{\rm sub}$ is given by
\begin{subequations}%
\label{eq:alphas-renorm}%
\begin{eqnarray}
\bar J_R^{\rm sub}=
[\bm{1}
+\alpha_s Z_{\bar J^{\rm sub}}^{(1)}+O(\alpha_s^2)]\otimes
\bar J^{\rm sub},
\end{eqnarray}
where
\begin{equation}
\label{eq:Z-bar-J-sub}%
Z_{\bar J^{\rm sub}}^{(1)}(1-x,p^2;\mu)= 
Z_{\bar J}^{(1)}(p^2;\mu) \delta(1-x)
-Z_{\bar S}^{(1)}[(1-x);\mu].
\end{equation}
\end{subequations}
 Equation (\ref{eq:alphas-renorm}) is the basis for our order-$\alpha_s$
 calculations of the renormalization of $\bar J^{\rm sub}$.

The $Z$ factors in Eq.~(\ref{eq:alphas-renorm}) are ultimately
    convolved with $\bar J^{\rm sub}$. [See
      Eq.~(\ref{eq:lowest-order-S-conv}).] In extracting the UV
divergences, we implement these convolutions (a simple multiplication in
the case of $Z_{\bar J}$) in order to keep track of the nonlocality of
the $Z_{\bar S}$ and in order to make use of the Dirac projector in
$\bar J^{\rm sub}$ to simplify expressions.

We compute the renormalization factor $Z_{\bar J}^{(1)}$ that is
associated with the unsubtracted radiative jet function $\bar J$
[Eq.~(\ref{defJF})] by evaluating the order-$\alpha_s$ UV divergence in
$\bar J$, namely ($Z_{\bar J}^{-1})^{(1)}$ and using ($Z_{\bar
  J}^{-1})^{(1)}=-Z_{\bar J}^{(1)}$. We convolve the result with $\bar
    J^{\rm sub}$.

We compute the renormalization factor $Z_{\bar S}^{(1)}$ that is
associated with the soft subtraction [Eq.~(\ref{eq:softsub})] by
evaluating $[\bar S^{(0)}_R\otimes (Z_{\bar S}^{-1})^{(1)}] \otimes
\bar J^{\rm sub}$.  Then, we make use of
Eq.~(\ref{eq:lowest-order-S-conv}) to eliminate the trivial convolution
involving $\bar S_R^{(0)}$, and we use $(Z_{\bar S}^{-1})^{(1)}=-Z_{\bar
  S}^{(1)}$. Finally, we combine this result with the result for
$(Z_{\bar J}^{-1})^{(1)}\otimes \bar J^{\rm sub}$ according to
Eq.~(\ref{eq:alphas-renorm}).

We evaluate many of the integrals by using contour integration. In these
evaluations, we take, for definiteness, $p_+>0$.  This choice
    corresponds to a time-like argument $p^2>0$ in the radiative jet
    function because $p^2\approx p_+p_-$ and $p_-\approx k_{1-}$ is
    always positive. However, our result for the anomalous dimension of
    the radiative jet function in the time-like case can be continued
    analytically to obtain the anomalous dimension in the
    space-like ($p^2<0$) case.

We carry out the computation in the Feynman gauge. 

\begin{figure}
\begin{center}
\includegraphics[width=1.0\columnwidth]{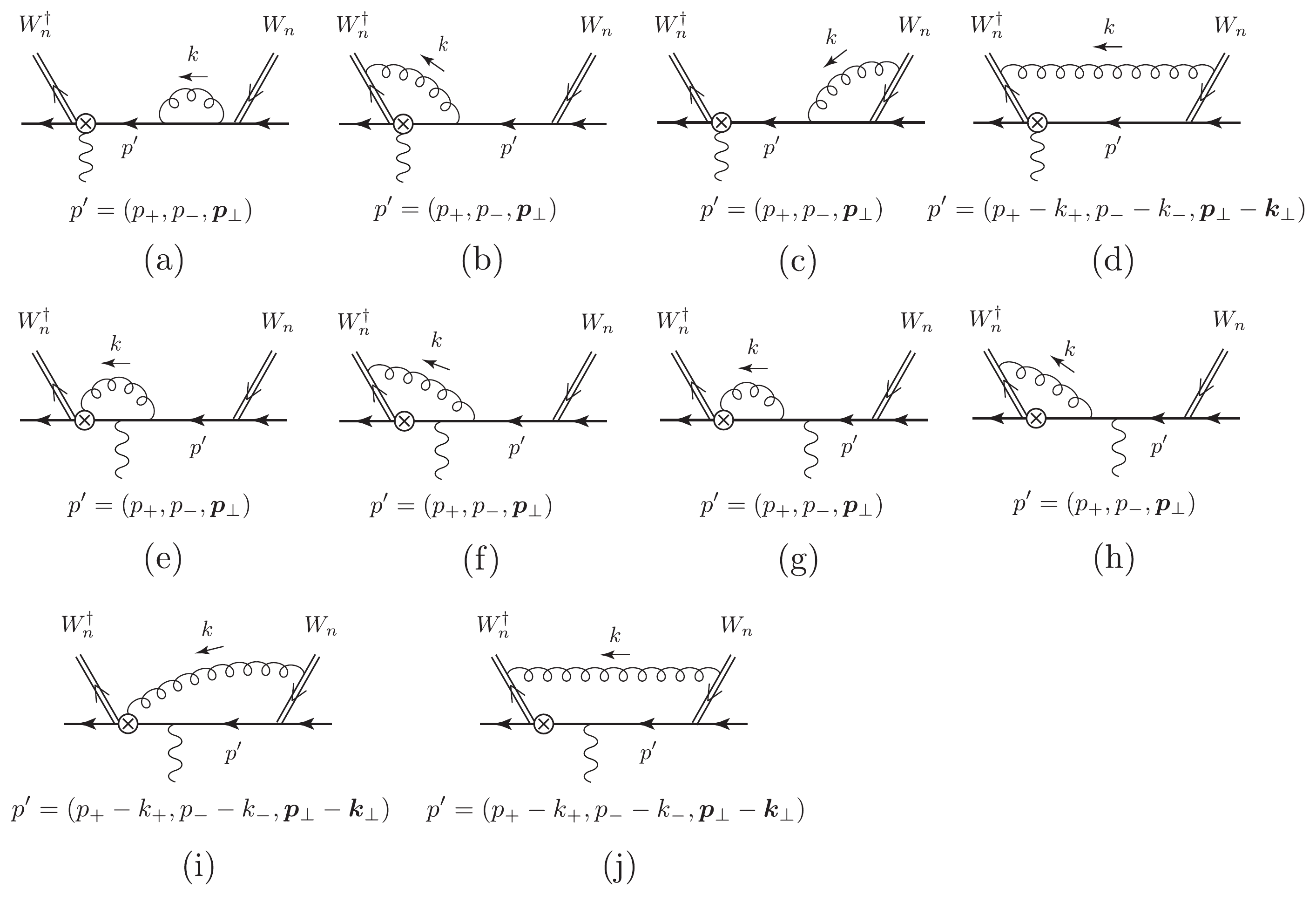}
\caption{Feynman diagrams that contribute order-$\alpha_s$ corrections
  to the radiative jet function in the Feynman gauge.
  The external momentum $p'$ of the
  heavy-quark propagator is shown below each diagram. The diagrams in
  the first row arise from $\bar J_{A}$, and the diagrams in the second
  and third rows arise from $\bar J_{G}$.  The crossed circle
  corresponds to the covariant derivative in Eq.~(\ref{eq:softsub}). Our
  graphical notation differs from that in Ref.~\cite{Liu:2019oav}, in
  which the crossed circle represents both the covariant derivative and
  Wilson lines.
\label{fig:jet} }
\end{center}
\end{figure}

\begin{figure}
\begin{center}
\includegraphics[width=1.0\columnwidth]{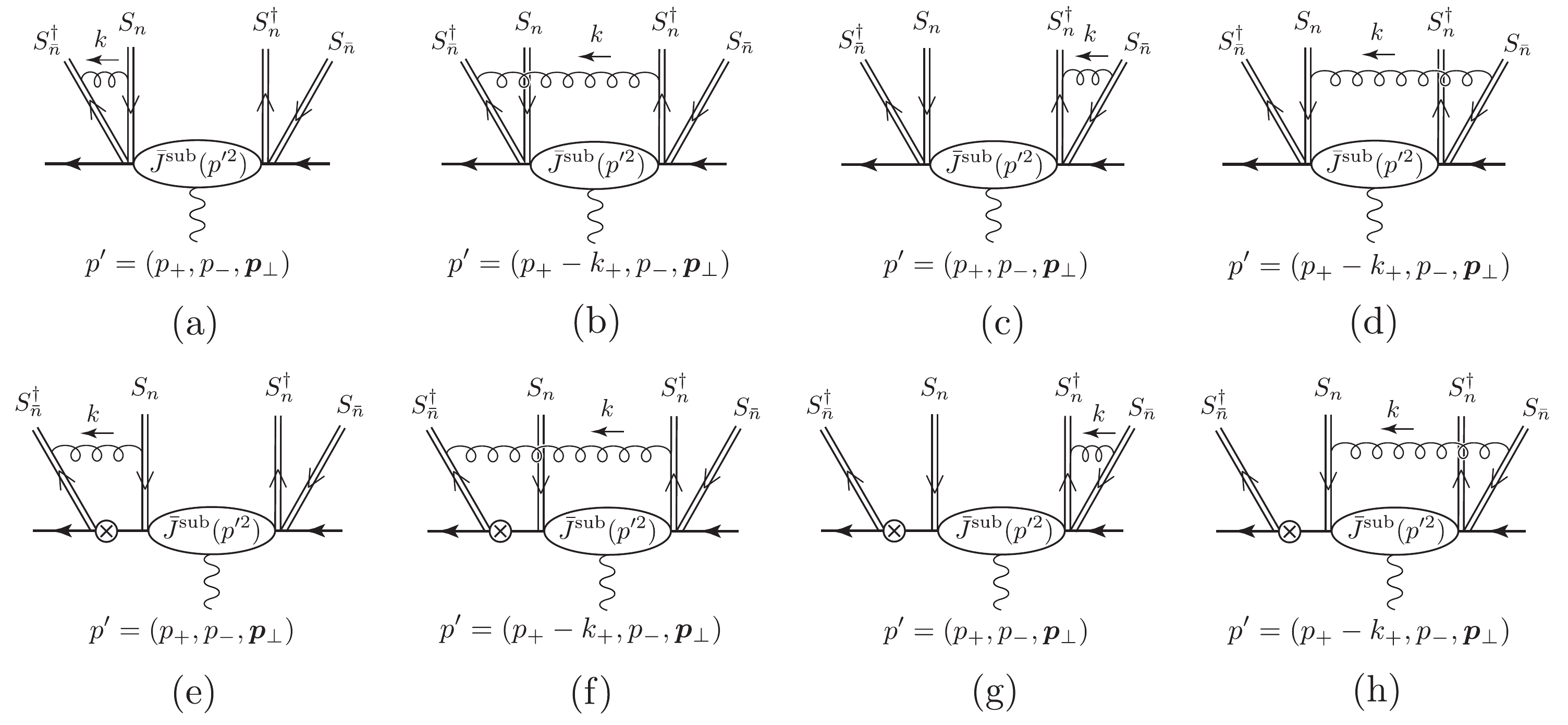}
\caption{Feynman diagrams that contribute to the renormalization of the
  soft-subtraction function $\bar S$ in order $\alpha_s$ in the
  Feynman gauge.
  The blobs in
  the diagrams represent $\bar J^{\rm sub}(p'^2)$, and the corresponding
  $p'$ is shown below each diagram.  The diagrams (g) and (h) do not
  contribute, according to the definition of the soft-subtraction
  function that is given in Sec.~\ref{sec:radiative-jet-function},
  because no soft gluons attach either to the covariant derivative or to
  $S_{\bar n}^\dagger$.  
  \label{fig:soft-sub}
}
\end{center}
\end{figure}

Feynman diagrams that potentially contribute in order $\alpha_s$ to the
renormalization of the radiative jet function $\bar J$ in the Feynman
gauge are shown in Fig.~\ref{fig:jet}. Feynman diagrams that potentially
contribute in order $\alpha_s$ to the renormalization of the soft
subtractions in the Feynman gauge are shown in
Fig.~\ref{fig:soft-sub}. In Fig.~\ref{fig:soft-sub}, the blobs represent
$\bar J^\textrm{sub}$,
which, as we have mentioned, we make explicit in order to keep
track of the nonlocal nature of $Z_{\bar S}$ and the Dirac algebra.

In carrying out these computations we make use of the Feynman rules for
the Wilson lines, which can be obtained from the definitions in 
Eqs.~(\ref{eq:collinear-wilson})
and (\ref{def:Wilson-line}). A $W_n$ vertex contributes a factor
$ig_s T^a\bar{n}^\mu$. A $W_n$ propagator contributes a factor
\begin{eqnarray}
\label{eq:feynman-collinear-wilson}
\frac{i}{\pm\bar{n}\cdot k+i\varepsilon}.
\end{eqnarray}
The plus sign applies when $k$ flows into the Wilson line, and the minus
    sign applies when $k$ flows out of the Wilson line. In the case of
    $W_{\bar n}$, $\bar n$ is replaced with $n$.  The Feynman rules for
    $W_n^\dagger$ and $W_{\bar n}^\dagger$ can be obtained from those for
    $W_n$ and $W_{\bar n}$, respectively, by taking the Hermitian
    conjugate and reversing the sign of the momentum in the
    propagator. The Feynman rules for $S_n$ ($S_{\bar n}$) are identical
    to the Feynman rules for $W_{\bar n}$ ($W_n$).


%%%%%%%%%%%%%%%%%%%%%%%%%%%%%%%%%%%%%%%%%%%%%%%%%%%%%%%%%%%%%%%%%%%%%%%%%%%%%%%%%%%%%%%%%%%%%%%%%%
\subsection{UV divergences of
  $\pmb{\bar{J}}^{\textbf{sub}}_\textbf{\!\textit{A}}$(\textbf{\textit{p}}$^{\textbf{2}}$)
  at order $\pmb{\alpha_s}$
\label{sec:J-A-alphas}}
Let us discuss the contributions to $\bar{J}^{\rm sub}_{A}(p^{2})$
first. The diagrams that contribute to $\bar J$ in order $\alpha_s$ are
shown in Fig.~\ref{fig:jet}.  Among the diagrams in Fig.~\ref{fig:jet},
diagrams (a)--(d) are contributions to $\bar{J}_{A}(p^{2})$.  

Diagram (d), which could potentially yield a nonlocal renormalization,
    has a UV-convergent power count.  This is an example of the general
    argument that we have given in Sec.~\ref{sec:renorm} that $\bar J$
    has only local renormalizations. Diagram (d) also gives a
    vanishing contribution, since the collinear gluon is connected to
    two collinear Wilson lines that are in the same direction, which
    leads to both a vanishing numerator and a vanishing $k_{+}$ contour
    integral.

Diagram (a) in Fig.~\ref{fig:jet} is simply
the quark self-energy diagram, and the corresponding amplitude is given
by
\begin{eqnarray}
\label{eq:jet-diagram-a}%
i\mathcal{A}_{ \textrm{(a)}} &=&
ie_{q}g_{s}^{2}C_{F}\left(\frac{\mu^{2}e^{\gamma_{E}}}{4\pi}\right)^{\epsilon}\int
\frac{d^{D} k}{(2\pi)^{D}}
\frac{\slashed{\varepsilon}^{*}_{\perp}
P_n\slashed{p}
\gamma^{\mu}(\slashed{p}-\slashed{k})\gamma_{\mu}\slashed{p}
P_{\bar n}}{
      (p^2+i\varepsilon)^2
  (k^{2}+i\varepsilon)[(p-k)^{2}+i\varepsilon]}\nn\\ 
  &=&
-e_{q}g_{s}^{2}C_{F}\left(\frac{\mu^{2}e^{\gamma_{E}}}{4\pi}\right)^{\epsilon}\slashed{\varepsilon}^{*}_{\perp}\frac{\slashed{n}}{2}
\frac{i\bar{n}\cdot p}{p^2+i\varepsilon}
\int \frac{d^{D}
  k}{(2\pi)^{D}}
\frac{(D-2)\left(1-\frac{n\cdot k}{\bar{n}\cdot p}\right)}
{(k^{2}+i\varepsilon)[(p-k)^{2}+i\varepsilon]},
\end{eqnarray}
where $C_F=(N_c^2-1)/(2N_c)$ and $N_c$ is the number of colors.

There is no soft subtraction in Fig.~\ref{fig:soft-sub} that corresponds
to this contribution. This follows from the fact that, in
Eq.~(\ref{eq:jet-diagram-a}), $(p-k)^{2}+i\varepsilon \simeq p^{2}-
p_{-}k_{+}+i\varepsilon$ when $k$ is soft, which implies that there is
no pinch that prevents the $k_{-}$ integration contour from being
deformed out of the soft region.  From the Dirac structure in
Eq.~(\ref{eq:jet-diagram-a}), we can see that the associated UV
divergence corresponds to a quark wave-function
renormalization. Extracting the UV divergence and convolving
(multiplying) with $\bar J^{\rm sub}$, we obtain
\begin{eqnarray}
(-Z_{\bar J}^{(1)}\otimes \bar  J^{\rm sub})_{(\textrm{a})}
= - \frac{\alpha_{s}C_{F}}{4\pi}\frac{1}{\epsilon_{\textrm{UV}}}\bar{J}^{\textrm{sub}}(p^{2}).
\end{eqnarray}
The subscript UV indicates that the origin of the divergence is UV.

%Therefore, we have
%\begin{eqnarray}
%\bar{J}_{A,2}^{\textrm{sub},\textrm{UV}}(p^{2})= - \frac{\alpha_{s}C_{F}}{4\pi}\frac{1}{\epsilon_{\textrm{UV}}}\bar{J}(p^{2}).
%\end{eqnarray}
The amplitudes of diagrams (b) and (c) in Fig.~\ref{fig:jet} are given by
\begin{eqnarray}
\label{eq:diagrams-b-c}
i\mathcal{A}_{ \textrm{(b)}}
&=& 
-ie_{q}g_{s}^{2}C_{F}\left(\frac{\mu^{2}e^{\gamma_{E}}}{4\pi}\right)^{\epsilon}\int \frac{d^{D} k}{(2\pi)^{D}} \frac{\slashed{\varepsilon}^{*}_{\perp}
P_n
(\slashed{p}-\slashed{k})\slashed{\bar{n}}\slashed{p}
P_{\bar n}}
{(p^2+i\varepsilon)(k^{2}+i\varepsilon)[(p-k)^{2}+i\varepsilon]
(\bar{n}\cdot k+i\varepsilon)}
\nn\\
&=&  -2e_{q}g_{s}^{2}C_{F}\left(\frac{\mu^{2}e^{\gamma_{E}}}{4\pi}\right)^{\epsilon}\slashed{\varepsilon}^{*}_{\perp}\frac{\slashed{n}}{2}\frac{i\bar{n}\cdot p}{p^2+i\varepsilon}
\int \frac{d^{D} k}{(2\pi)^{D}}\frac{\bar{n}\cdot(p-k)}
{(k^{2}+i\varepsilon)[(p-k)^{2}+i\varepsilon]
(\bar{n}\cdot k+i\varepsilon)},
\nn\\
i\mathcal{A}_{\textrm{(c)}}&=& 
ie_{q}g_{s}^{2}C_{F}\left(\frac{\mu^{2}e^{\gamma_{E}}}{4\pi}\right)^{\epsilon}
\int \frac{d^{D} k}{(2\pi)^{D}} 
\frac{\slashed{\varepsilon}^{*}_{\perp}
P_n
\slashed{p}\slashed{\bar{n}}(\slashed{p}-\slashed{k})
P_{\bar n}}
{(p^{2}+i\varepsilon)(k^{2}+i\varepsilon)[(p-k)^{2}+i\varepsilon]
(-\bar{n}\cdot k+i\varepsilon)}\nn\\
&=&  2e_{q}g_{s}^{2}C_{F}\left(\frac{\mu^{2}e^{\gamma_{E}}}{4\pi}\right)^{\epsilon}\slashed{\varepsilon}^{*}_{\perp}\frac{\slashed{n}}{2}\frac{i\bar{n}\cdot p}{p^2+i\varepsilon}\int \frac{d^{D} k}{(2\pi)^{D}}\frac{\bar{n}\cdot(p-k)}{(k^{2}+i\varepsilon)
[(p-k)^{2}+i\varepsilon](-\bar{n}\cdot k+i\varepsilon)}.
\phantom{XX}
\end{eqnarray}
We convolve
(multiply) the UV-divergent contribution 
in Eq.~(\ref{eq:diagrams-b-c})
with $\bar J^{\rm sub}$ to obtain
\begin{eqnarray}
\label{eq:UV-diagrams-b-c}%
\alpha_s(-Z_{\bar J}^{(1)}\otimes \bar  J^{\rm sub})_{(\textrm{b})}
&=&  -2ig_{s}^{2}C_{F}\left(\frac{\mu^{2}e^{\gamma_{E}}}{4\pi}\right)^{\epsilon}
\left[
\int \frac{d^{D} k}{(2\pi)^{D}}\frac{\bar n \cdot (k-p)
\bar{J}^{\textrm{sub}}(p^{2})}
{(k^{2}+i\varepsilon)[(p-k)^{2}+i\varepsilon]
(\bar n\cdot k+i\varepsilon)}
\right]_\textrm{UV},
\nonumber \\
\alpha_s(-Z_{\bar J}^{(1)}\otimes \bar  J^{\rm sub})_{(\textrm{c})}
&=&  2ig_{s}^{2}C_{F}\left(\frac{\mu^{2}e^{\gamma_{E}}}{4\pi}\right)^{\epsilon}
\left[
\int \frac{d^{D} k}{(2\pi)^{D}}\frac{\bar n \cdot(k-p)
\bar{J}^{\textrm{sub}}(p^{2})}{(k^{2}+i\varepsilon)
[(p-k)^{2}+i\varepsilon]
(-\bar n\cdot k+i\varepsilon)}
\right]_\textrm{UV}.
\phantom{XX}
\end{eqnarray}
Here, the subscript UV indicates that only the UV-divergent part of the
expression is to be kept.

As we will explain, the soft subtractions that correspond to diagram (b)
in Fig.~\ref{fig:jet} are given by diagrams (a) and (b) in
Fig.~\ref{fig:soft-sub}, and the soft subtractions that correspond to
diagram (c) in Fig.~\ref{fig:jet} are given by diagrams (c) and (d) in
Fig.~\ref{fig:soft-sub}.  After working out the numerator algebra, we
obtain the following contributions:
\begin{eqnarray}
\label{eq:UV-soft-a+b}%
\alpha_s(-Z_{\bar S}^{(1)}\otimes \bar 
J^{\rm sub})_{(\textrm{a+b})}
&=&
-in\cdot \bar ng_{s}^{2}C_{F}\left(\frac{\mu^{2}e^{\gamma_{E}}}{4\pi}\right)^{\epsilon}
\left[
\int \frac{d^{D}
  k}{(2\pi)^{D}}\frac{\bar{J}^{\textrm{sub}}(p^{2})
-\bar{J}^{\textrm{sub}}
\left( p^{2}-\frac{2(\bar n \cdot p) (n\cdot k)}
{n\cdot \bar n}\right)}
{(k^{2}+i\varepsilon)(n\cdot k-i\varepsilon)
(\bar n\cdot k+i\varepsilon)}
\right]_\textrm{UV},
\nonumber \\
\alpha_s(-Z_{\bar S}^{(1)}\otimes \bar 
J^{\rm sub})_{(\textrm{c+d})}
&=&
-in\cdot \bar ng_{s}^{2}C_{F}\left(\frac{\mu^{2}e^{\gamma_{E}}}{4\pi}\right)^{\epsilon}
\left[
\int \frac{d^{D}
  k}{(2\pi)^{D}}\frac{\bar{J}^{\textrm{sub}}(p^{2})
-\bar{J}^{\textrm{sub}}\left(p^{2}-\frac{2(\bar n \cdot p) (n\cdot k)}
{n\cdot \bar n}\right)}{(k^{2}+i\varepsilon)
(n\cdot k+i\varepsilon)(\bar n\cdot k-i\varepsilon)}
\right]_\textrm{UV}.
\nonumber \\
\end{eqnarray}

We can see, in the case of the lowest-order expression for $\bar
J^\textrm{sub}$ in Eq.~(\ref{eq:j-bar-sub-0}), that the expression on
the right side of Eq.~(\ref{eq:UV-soft-a+b}) is equal to the
soft-approximation of the expression on the right side of
Eq.~(\ref{eq:UV-diagrams-b-c}). This is the sense in which these soft
    subtractions correspond to diagram (b) in Fig.~\ref{fig:jet}.  
    Hence, in the
    case of the lowest-order expression for $\bar J^\textrm{sub}$, the
    IR divergences in Eq.~(\ref{eq:UV-soft-a+b}) cancel those in
    Eq.~(\ref{eq:UV-diagrams-b-c}). However, this cancellation does not
    hold in general because Eq.~(\ref{eq:UV-diagrams-b-c}) is missing some
    of the jet-function contributions whose IR divergences would be
    canceled by the soft-subtraction contribution in
    Eq.~(\ref{eq:UV-soft-a+b}). Recall that the expression in
    Eq.~(\ref{eq:UV-soft-a+b}) contains the contributions of many
    soft-subtraction subgraphs that are summed through the use of the
    graphical Ward identities, while the expression in
    Eq.~(\ref{eq:UV-diagrams-b-c}) contains only the contributions from
    those subgraphs that yield the UV divergences that are associated
    with the radiative-jet operator matrix element.

We can see as well, in the case of the lowest-order expression for $\bar
    J^\textrm{sub}$, that the nonlocal contributions on the right side
    of Eq.~(\ref{eq:UV-soft-a+b}) vanish. However, this feature is also
    special to the lowest-order case.

Note that, if one carries out the integration over $k_-$ in
    Eq.~(\ref{eq:UV-soft-a+b}) by contour integration, the resulting
    expression contains a scaleless integral in $k_\perp$ that yields a
    factor $1/\epsilon_{\rm UV}-1/\epsilon_{\rm IR}$.  Therefore, if one
    does not distinguish between UV and IR poles in dimensional
    regularization, the soft subtraction vanishes. Hence, its role
    is to convert nonlocal IR poles to nonlocal UV poles.

Owing to the presence of the Wilson-line denominators $k_-\pm i\ve$, the
expressions in Eqs.~(\ref{eq:UV-diagrams-b-c}) and
(\ref{eq:UV-soft-a+b}) develop rapidity divergences as $k_+/k_-\to
\infty$.  These rapidity divergences cancel in the difference between
Eqs.~(\ref{eq:UV-diagrams-b-c}) and (\ref{eq:UV-soft-a+b}). Therefore,
in order to avoid introducing a rapidity regulator, we compute the
expressions in Eqs.~(\ref{eq:UV-diagrams-b-c}) and
(\ref{eq:UV-soft-a+b}) together.  Reparameterization invariance
\cite{Manohar:2002fd} (separate rescaling of $n$ and $\bar n$)
guarantees that, aside from the overall kinematic factor $\bar{n}\cdot
    p$ [see Eq.~(\ref{defJF})], the expressions in
    Eqs.~(\ref{eq:UV-diagrams-b-c}) and (\ref{eq:UV-soft-a+b}) are
    functions only of $p^2$ or $\frac{(n\cdot p) (\bar n\cdot
      p)}{n\cdot\bar n}\approx p^2$.  (We will see this explicitly
    below.)  Therefore, once we have extracted the $k$-independent
        factor $\bar n\cdot p$, the remainder of the integrand is
        invariant under $p\to -p$, and we can make the variable change
        $k\to -k$ to find that
\begin{equation}
(-Z_{\bar J}^{(1)}\otimes \bar  J^{\rm sub})_{(\textrm{b})}
-
(-Z_{\bar S}^{(1)}\otimes \bar 
J^{\rm sub})_{(\textrm{a+b})}
=
(-Z_{\bar J}^{(1)}\otimes \bar  J^{\rm sub})_{(\textrm{c})}
-
(-Z_{\bar S}^{(1)}\otimes \bar 
J^{\rm sub})_{(\textrm{c+d})}.
\end{equation}
Therefore, in the analysis of these contributions, we compute only the
    combination $(-Z_{\bar J}^{(1)}\otimes \bar J^{\rm
  sub})_{(\textrm{b})} - (-Z_{\bar S}^{(1)}\otimes \bar J^{\rm
  sub})_{(\textrm{a+b})}$. 

We extract the contribution 
\begin{eqnarray}
I_{\textrm{b}}(p^{2})= -2ig_{s}^{2}C_{F}\left(\frac{\mu^{2}e^{\gamma_{E}}}{4\pi}\right)^{\epsilon}\int \frac{d^{D} k}{(2\pi)^{D}}
\frac{\bar{J}^{\textrm{sub}}(p^{2})}
{(k^{2}+i\varepsilon)[(p-k)^{2}+i\varepsilon]}
\end{eqnarray}  
from $(-Z_{\bar J}^{(1)}\otimes \bar  J^{\rm sub})_{(\textrm{b})}$, which gives the UV pole
\begin{eqnarray}
I_{\textrm{b}}^{\textrm{UV}}(p^{2}) =  \frac{\alpha_{s}C_{F}}{2\pi}\frac{1}{\epsilon_{\textrm{UV}}}\bar{J}^{\textrm{sub}}(p^{2}).
\end{eqnarray}
Then, we combine the remaining part of the integrals in
$(-Z_{\bar J}^{(1)}\otimes \bar  J^{\rm sub})_{(\textrm{b})}$
and the soft subtractions
$(-Z_{\bar S}^{(1)}\otimes \bar 
J^{\rm sub})_{(\textrm{a+b})}$ to obtain
\begin{eqnarray}
&&
I_{\textrm{sub}}(p^{2})
\nonumber \\
&=&  -2ig_{s}^{2}C_{F}\left(\frac{\mu^{2}e^{\gamma_{E}}}{4\pi}\right)^{\epsilon}\int \frac{d^{D} k}{(2\pi)^{D}}
\bigg\{\frac{-p_{-}\bar{J}^{\textrm{sub}}(p^{2})}
{(k^{2}+i\varepsilon)[(p-k)^{2}+i\varepsilon](k_{-}+ i\varepsilon)}-\frac{\bar{J}^{\textrm{sub}}(p^{2})-\bar{J}^{\textrm{sub}}(p^{2}-p_{-}k_{+})}{(k^{2}+i\varepsilon)(k_{+}-i\varepsilon)(k_{-}+i\varepsilon)}\bigg\}
\nn\\
&=& -2ig_{s}^{2}C_{F}\left(\frac{\mu^{2}e^{\gamma_{E}}}{4\pi}\right)^{\epsilon}\int \frac{d^{D} k}{(2\pi)^{D}}
\Bigg\{\frac{-p_{-}\bar{J}^{\textrm{sub}}(p^{2})}
{(k^{2}+i\varepsilon)(k_{-}+ i\varepsilon)}\bigg[\frac{1}
{(p-k)^{2}+i\varepsilon}-\frac{1}{p^{2}-p_{-}k_{+}+i\varepsilon}\bigg]
\nn\\ &&
\quad\quad\quad\quad\quad\quad\quad\quad
+\frac{1}{(k^{2}+i\varepsilon)(k_{+}-i\varepsilon)(k_{-}+i\varepsilon)}
\left[
\bar{J}^{\textrm{sub}}(p^{2}-p_-k_+)
-
\frac{p^2\bar{J}^{\textrm{sub}}(p^{2})}{p^2-p_-k_++i\varepsilon}
\right]
\Bigg\}.
\end{eqnarray}

Carrying out the $k_{-}$ contour integration by deforming the contour
    into the lower half-plane (making the assumption $p_+>0$) and
using $p_+p_-\approx p^2$, we find that
\begin{eqnarray}
&&
I_{\textrm{sub}}(p^{2})
\nn\\
&=& g_{s}^{2}C_{F}\left(\frac{\mu^{2}e^{\gamma_{E}}}{4\pi}\right)^{\epsilon}
\int\frac{d^{D-2} k_{\perp}}{(2\pi)^{D-2}}
\frac{1}{-\bm{k}_\perp^2}
\nonumber \\
&&
\times
\int \frac{dk_+}{2\pi}
\Bigg\{
\left[
\frac{\theta(p_+-k_+)}{p^{2}-p_{-}k_{+}-\bs{k}_{\perp}^{2}+i\ve}
-
\frac{\theta(-k_+)}{p^{2}-p_{-}k_{+}+i\varepsilon}
-
\frac{\theta(k_+)\theta(p_+-k_+)}
{p^{2}-p_{-}k_{+}-\frac{p_{+}}{k_{+}}\bs{k}_{\perp}^{2}+i\ve}
\right]
p_{-}\bar{J}^{\textrm{sub}}(p^{2})
\nn\\ &&
\quad\quad\quad\quad\quad
-\frac{\theta(-k_+)}
{k_{+}-i\varepsilon}
\left[
\bar{J}^{\textrm{sub}}(p^{2}-p_-k_+)
-\frac{p^2\bar{J}^{\textrm{sub}}(p^{2})}{p^2-p_-k_++i\varepsilon}
\right]
\Bigg\}.
\end{eqnarray}
Rewriting the $\theta$ function in the first term according to
$\theta(p_+-k_+)=\theta(p_+-k_+)\theta(k_+)+\theta(-k_+)$, we arrive at
\begin{eqnarray}
&&
I_{\textrm{sub}}(p^{2})
\nn\\
&=& g_{s}^{2}C_{F}\left(\frac{\mu^{2}e^{\gamma_{E}}}{4\pi}\right)^{\epsilon}
\int\frac{d^{D-2} k_{\perp}}{(2\pi)^{D-2}}
\frac{1}{-\bm{k}_\perp^2}
\nonumber \\
&&
\times
\int \frac{dk_+}{2\pi}
\Bigg\{
\theta(-k_+)\left(
\frac{1}{p^{2}-p_{-}k_{+}-\bs{k}_{\perp}^{2}+i\ve}
-
\frac{1}{p^{2}-p_{-}k_{+}+i\varepsilon}
\right)
p_{-}\bar{J}^{\textrm{sub}}(p^{2})
\nonumber \\
&&
\quad\quad\quad\quad
+
\theta(k_+)\theta(p_+-k_+)
\left(
\frac{1}{p^{2}-p_{-}k_{+}-\bs{k}_{\perp}^{2}+i\ve}
-
\frac{1}
{p^{2}-p_{-}k_{+}-\frac{p_{+}}{k_{+}}\bs{k}_{\perp}^{2}+i\ve}
\right)
p_{-}\bar{J}^{\textrm{sub}}(p^{2})
\nn\\ &&
\quad\quad\quad\quad
-\frac{\theta(-k_+)}
{k_{+}-i\varepsilon}
\left[
\bar{J}^{\textrm{sub}}(p^{2}-p_-k_+)
-\frac{p^2\bar{J}^{\textrm{sub}}(p^{2})}{p^2-p_-k_++i\varepsilon}
\right]
\Bigg\}.
\end{eqnarray}

After carrying out the $k_{\perp}$ integration, we obtain
\begin{eqnarray}
&&
I_{\textrm{sub}}(p^{2})
\nonumber \\
&=&
\frac{g_{s}^{2}C_{F}}{4\pi}\Bigg\{
\left(\mu^{2}e^{\gamma_{E}}\right)^{\epsilon}\Gamma(\e_{\textrm{UV}})
\bigg[
\int_{-\infty}^{0}\frac{d k_{+}}{2\pi}
\frac{-p_{-}}{(-p^{2}+p_{-}k_{+}-i\ve)^{1+\e}}
+
\int_{0}^{p_+}
\frac{d k_{+}}{2\pi}\frac{-p_{-}(1-(\frac{p_{+}}{k_{+}})^{\e})}{(-p^{2}+
p_{-}k_{+}-i\ve)^{1+\e}}
\bigg]\bar{J}^{\textrm{sub}}(p^{2})\nn\\
&&
\quad\quad\quad
+\left(\frac{1}{\e_{\textrm{UV}}}-\frac{1}{\e_{\textrm{IR}}}\right)
\int_{-\infty}^{0} \frac{d k_{+}}{2\pi}\frac{1}{k_{+}}
\left[
\bar{J}^{\textrm{sub}}(p^{2}-p_{-}k_{+})
-
\frac{p^{2}\bar{J}^{\textrm{sub}}(p^{2})}{p^{2}-p_{-}k_{+}+i\ve}\right]
\Bigg\}.
\end{eqnarray}
Making the change of variables $k_+=(1-x)p_+$ and using $p_+p_-\approx
    p^2$, we have
\begin{eqnarray}
&&
I_{\textrm{sub}}(p^{2})
\nonumber \\
&=&\frac{\alpha_{s}C_{F}}{2\pi}\Bigg\{\left(\frac{\mu^{2}e^{\gamma_{E}}}{-p^{2}-i\varepsilon}\right)^{\epsilon}\Gamma(\e_{\textrm{UV}})\bigg[\int_{1}^{\infty}dx\frac{1}{x^{1+\e}}+\int_{0}^{1}dx\frac{1-(1-x)^{-\e}}{x^{1+\e}}\bigg]\bar{J}^{\textrm{sub}}(p^{2})\nn\\
&&
\quad\quad\quad
-\left(\frac{1}{\e_{\textrm{UV}}}-\frac{1}{\e_{\textrm{IR}}}\right)\int_{1}^{\infty}dx
\frac{1}{x(x-1)}\left[x\bar{J}^{\textrm{sub}}(xp^{2})-\bar{J}^{\textrm{sub}}(p^{2})\right]\Bigg\}\nn\\
&=&\frac{\alpha_{s}C_{F}}{2\pi}\Bigg\{\left(\frac{\mu^{2}e^{\gamma_{E}}}{-p^{2}-i\varepsilon}\right)^{\epsilon}\frac{\Gamma(-\e_{\textrm{UV}})^{2}\Gamma(1+\e)}{\Gamma(1-2\e)}\bar{J}^{\textrm{sub}}(p^{2})
-\left(\frac{1}{\e_{\textrm{UV}}}-\frac{1}{\e_{\textrm{IR}}}\right)\int_{1}^{\infty}dx
\frac{x\bar{J}^{\textrm{sub}}(xp^{2})}{[x(x-1)]_{+}}\Bigg\},
\nonumber \\
\end{eqnarray}
where we have used $g_s^2=4\pi \alpha_s$, and the plus distribution
is defined by
\begin{eqnarray}
\int_{0}^{\infty}dx\frac{f(x)}{[g(x)]_{+}} = \int_{0}^{\infty}\frac{f(x)-f(1)}{[g(x)]_{+}}.
\end{eqnarray}
Now we can extract the UV divergences of $I_{\textrm{sub}}(p^{2})$:
\begin{eqnarray}
I_{\textrm{sub}}^{\textrm{UV}}(p^{2}) &= & \frac{\alpha_{s}C_{F}}{2\pi}\Bigg\{\left[\frac{1}{\e^{2}_{\textrm{UV}}}+\frac{1}{\e_{\textrm{UV}}}\ln\left(\frac{\mu^{2}}{-p^{2}-i\varepsilon}\right)\right]\bar{J}^{\textrm{sub}}(p^{2})-\frac{1}{\e_{\textrm{UV}}}\int_{1}^{\infty}
dx \frac{x\bar{J}^{\textrm{sub}}(xp^{2})}{[x(x-1)]_{+}}\Bigg\}.
\phantom{XX}
\end{eqnarray}
Using Eq.~(\ref{eq:alphas-renorm}), we find that the total of the UV
    divergences in $\bar{J}^{\textrm{sub}}_{A}(p^{2})$ is
\begin{eqnarray}
\label{eq:JAsub}
&&
\alpha_s
(-Z_{\bar J^{\rm sub}}^{(1)}\otimes 
\bar J^{\rm sub})_{A}
\nonumber \\
&=& 
\alpha_s
(-Z_{\bar J}^{(1)}\otimes \bar  J^{\rm sub})_{(\textrm{a})}
+2\left[\alpha_s(-Z_{\bar J}^{(1)}\otimes \bar  J^{\rm sub})_{(\textrm{b})}
-
\alpha_s
(-Z_{\bar S}^{(1)}\otimes \bar 
J^{\rm sub})_{(\textrm{a+b})}\right]
\nonumber \\
&=&
- \frac{\alpha_{s}C_{F}}{4\pi}\frac{1}{\epsilon_{\textrm{UV}}}\bar{J}^{\textrm{sub}}(p^{2})
+2\left[I_{\textrm{b}}^{\textrm{UV}}(p^{2})+ I_{\textrm{sub}}^{\textrm{UV}}(p^{2})\right]
\nn\\
&=&
\frac{\alpha_{s}C_{F}}{4\pi}\Bigg\{\left[\frac{4}{\e^{2}_{\textrm{UV}}}+\frac{4}{\e_{\textrm{UV}}}\ln\left(\frac{\mu^{2}}{-p^{2}-i\varepsilon}\right)+\frac{3}{\e_{\textrm{UV}}}\right]\bar{J}^{\textrm{sub}}(p^{2})-\frac{4}{\e_{\textrm{UV}}}\int_{1}^{\infty}
dx \frac{x\bar{J}^{\textrm{sub}}(xp^{2})}{[x(x-1)]_{+}}\Bigg\}.
\phantom{XX}
\end{eqnarray}

%%%%%%%%%%%%%%%%%%%%%%%%%%%%%%%%%%%%%%%%%%%%%%%
%%%%%%%%%%%%%%%%%%%%%%%%%%%%%%%%%%%%%%%%%%%%%%%
\subsection{UV divergences of $\pmb{\bar{J}}^{\textbf{sub}}_\textbf{\!\textit{G}}$(\textbf{\textit{p}}$^{\textbf{2}}$) at order $\pmb{\alpha_s}$\label{sec:J-G-alphas}}
Among the diagrams in Fig.~\ref{fig:jet}, diagrams (e)--(i) are
contributions to $\bar{J}_{G}(p^{2})$.  Only diagrams (e) and (f) give
non-vanishing contributions to $\bar{J}_{G}(p^{2})$.  
Diagrams (g) and (h) have the denominators $k^{2}+i\ve$, $\bar{n}\cdot
k+i\ve$, and $(k-\ell)^2+i\ve$, which lead to vanishing $k_{+}$ contour
integrals.  This vanishing of the contour integrals is an example of
the situation that we mentioned 
in Sec.~\ref{subsec:subtraction-factorization}: 
These diagrams are not
properly part of the radiative jet function because they are not
connected to the external photon by lines carrying
$n$-hard-collinear momenta and, hence, they do not yield pinch
singularities in the $n$-hard-collinear region.
 Diagrams (i) and (j), which
could potentially yield nonlocal renormalizations, have UV-convergent
power counts.  These are examples of the general argument that we have
given in Sec.~\ref{sec:renorm} that $\bar J$ has only local
renormalizations.  (Diagrams (i) and (j) also have a vanishing numerator
structure in the Feynman gauge.)

The combined amplitude of diagrams (e) and (f) in Fig.~\ref{fig:jet} is
given by
\begin{eqnarray}
\label{eq:AJ6}
i\mathcal{A}_{\textrm{(e+f)}} 
&=&
i e_{q}g_{s}^{2}C_{F}\left(\frac{\mu^{2}e^{\gamma_{E}}}{4\pi}\right)^{\epsilon}\int \frac{d^{D} k}{(2\pi)^{D}} 
\frac{\left(\gamma_{\mu}^{\perp}-\frac{\slashed{k}_{\perp}\bar{n}_{\mu}}
{\bar{n}\cdot k+i\varepsilon}\right)
P_n(\slashed{\ell}-\slashed{k})\slashed{\varepsilon}^{*}_{\perp}
(\slashed{p}-\slashed{k})\gamma^{\mu}\slashed{p}
P_{\bar n}}
{(k^{2}+i\varepsilon)[(p-k)^{2}+i\varepsilon]
[(\ell-k)^{2}+i\varepsilon](p^2+i\varepsilon)}
\nn\\
&=&e_{q}g_{s}^{2}C_{F}\left(\frac{\mu^{2}e^{\gamma_{E}}}{4\pi}\right)^{\epsilon}\slashed{\varepsilon}^{*}_{\perp}\frac{\slashed{n}}{2}
\frac{i\bar{n}\cdot p}{p^2+i\varepsilon}
\nonumber \\
&&
\times
\int \frac{d^{D} k}{(2\pi)^{D}}
\frac{-(D-6)p^{2}\frac{\bar{n}\cdot k}{\bar{n}\cdot p}-4k\cdot p+2(n\cdot k) (\bar{n}\cdot k)+k^{2}\left(D-6 +\frac{2\bar{n}\cdot p}
{\bar{n}\cdot k+i\varepsilon}\right)}
{(k^{2}+i\varepsilon)[(p-k)^{2}+i\varepsilon]
[(\frac{p^{2}}{2\bar{n}\cdot p} \bar{n}-k)^{2}+i\varepsilon]}.
\end{eqnarray}
We remind the reader that we retain only terms that contribute in
leading power in $\lambda$, under the assumption that the momentum
$\ell=p-k_1 \approx \frac{p^2}{\bar{n}\cdot p}\frac{\bar{n}}{2}$ is
soft and that the momenta $k$ and $k_1$ are $n$-hard collinear.

We extract the UV-divergent contribution from the amplitude in
    Eq.~(\ref{eq:AJ6}) and convolve it with $\bar J^{\rm sub}$ to
    obtain
\begin{eqnarray}
\label{eq:jet-e+f}%
\alpha_s(-Z_{\bar J}^{(1)}\otimes \bar  J^{\rm sub})_{(\textrm{e+f})}
 &=&-i g_{s}^{2}C_{F}\left(\frac{\mu^2 e^{\gamma_{E}}}{4\pi}\right)^{\epsilon}
\bigg\{
\int \frac{d^{D} k}{(2\pi)^{D}} 
\frac{\bar{J}^{\textrm{sub}}(p^{2})}
{(k^{2}+i\varepsilon)[(p-k)^{2}+i\varepsilon]
(k^{2}-p_{+}k_{-}+i\varepsilon)}
\nn\\
&&\times\left[-(D-4) p_{+}k_{-}+(D-4)k_{+}k_{-}+(6-D)\bs{k}^{2}_{\perp} 
-\frac{2\bs{k}^{2}_{\perp}p_{-}}{k_{-}+i\varepsilon}\right]
\bigg\}_\textrm{UV}.
\end{eqnarray}
 For convenience of computation, we split
 $(-Z_{\bar J}^{(1)}\otimes \bar  J^{\rm sub})_{(\textrm{e+f})}$ into two parts, one with the
 denominator $\frac{1}{k_{-} +i\varepsilon}$ and the other without it:
\begin{subequations}\label{eq:jet-e+f-split}%
\begin{eqnarray}
&&
\alpha_s(-Z_{\bar J}^{(1)}\otimes \bar  J^{\rm sub})_{(\textrm{e+f}),1}
\nonumber \\
&=& -i g_{s}^{2}C_{F}\left(\frac{\mu^{2}e^{\gamma_{E}}}{4\pi}\right)^{\epsilon}
\left[
\int \frac{d^{D} k}{(2\pi)^{D}} 
\frac{\left[2\e p_{+}k_{-}  -2\e k_{+}k_{-}+(2+2\e)\bs{k}^{2}_{\perp}\right]\bar{J}^{\textrm{sub}}(p^{2}) }
{(k^{2}+i\varepsilon)[(p-k)^{2}+i\varepsilon]
(k^{2}-p_{+}k_{-}+i\varepsilon)}
\right]_\textrm{UV},
\\[2ex]
&&
\alpha_s
(-Z_{\bar J}^{(1)}\otimes \bar  J^{\rm sub})_{(\textrm{e+f}),2}
\nonumber \\
&=& i g_{s}^{2}C_{F}\left(\frac{\mu^{2}e^{\gamma_{E}}}{4\pi}\right)^{\epsilon}
\left[
\int
\frac{d^{D} k}{(2\pi)^{D}} \frac{2\bs{k}^{2}_{\perp}
p_{-}\bar{J}^{\textrm{sub}}(p^{2})}
{(k_{-}+i\varepsilon)(k^{2}+i\varepsilon)
[(p-k)^{2}+i\varepsilon](k^{2}-p_{+}k_{-}+i\varepsilon)}
\right]_\textrm{UV}.\phantom{XXX}
\label{eq:jet-e+f-split-b}
\end{eqnarray}
\end{subequations}

The soft subtractions that correspond to 
diagrams (e) and (f) of Fig.~\ref{fig:jet}
are given by diagrams (e) and (f) of Fig.~\ref{fig:soft-sub}. Their
      UV-divergent contributions are
\begin{eqnarray}
\label{eq:soft-e+f}
&&
\alpha_s
(-Z_{\bar S}^{(1)}\otimes \bar J^{\rm sub})_{(\textrm{e+f})}
\nonumber \\
&=&
-2ig_s^2C_F
\left(\frac{\mu^{2}e^{\gamma_{E}}}{4\pi}\right)^{\epsilon}
\left[
\int \frac{d^Dk}{(2\pi)^D}
\frac{
P_n
k\sl_\perp 
\left(
\slashed{\ell}-\slashed{k}
\right)
P_{n}
\left\{
\bar{J}^{\textrm{sub}}(p^{2})
-\bar{J}^{\textrm{sub}}[(p-k_{+}\frac{\bar{n}}{2})^2]
\right\}}
{(k^2+i\varepsilon)
(k_+-i\varepsilon)
(k_-+i\varepsilon)
[(\ell-k)^2+i\varepsilon]}
\right]_\textrm{UV}
\nonumber \\
&=&i g_{s}^{2}C_{F}
\left(\frac{\mu^{2}e^{\gamma_{E}}}{4\pi}\right)^{\epsilon}
\left[
\int \frac{d^{D} k}{(2\pi)^{D}} \frac{-2\bs{k}^{2}_{\perp}\left[\bar{J}^{\textrm{sub}}(p^{2})-\bar{J}^{\textrm{sub}}(p^{2}-k_{+}p_{-})\right]}
{(k^{2}+i\varepsilon)(k_{-}+i\varepsilon)
(k_{+}-i\varepsilon)(k^{2}-p_{+}k_{-}+i\varepsilon)}
\right]_\textrm{UV}
\nonumber \\
&=&
i g_{s}^{2}C_{F}\left(\frac{\mu^{2}e^{\gamma_{E}}}{4\pi}\right)^{\epsilon}
\left[
\int \frac{d^{D} k}{(2\pi)^{D}} \frac{2\bs{k}^{2}_{\perp}p_{-}
\left[\bar{J}^{\textrm{sub}}(p^{2})
+ \frac{p^{2}\bar{J}^{\textrm{sub}}(p^{2})-(p^{2}-k_{+}p_{-})
\bar{J}^{\textrm{sub}}(p^{2}-k_{+}p_{-})}
{-k_{+}p_{-}+i\varepsilon}\right]
}
{(k^{2}+i\varepsilon)(k_{-}+i\varepsilon)
(p^{2}-k_{+}p_{-}+i\varepsilon)
(k^{2}-p_{+}k_{-}+i\varepsilon)}
\right]_\textrm{UV}.
\phantom{XX}
\end{eqnarray}
Here, we have used Eq.~(\ref{eq:ell-approx}) to drop terms that are
subleading in the scaling parameter $\lambda$ and have used the facts
    that $\ell_+=p_+$ and $P_n \bar{J}^{\rm sub}=\bar{J}^{\rm sub}$.
    As in our analysis in Sec.~\ref{sec:J-A-alphas}, we can see that, in
    the case of the lowest-order expression for $\bar J^\textrm{sub}$ in
    Eq.~(\ref{eq:j-bar-sub-0}), the expression in
    Eq.~(\ref{eq:soft-e+f}) is equal to the soft-approximation of the
    expression on the right side of Eq.~(\ref{eq:jet-e+f}).  This is the
    sense in which these soft subtractions correspond to the
    contributions of diagrams (e) and (f) in Fig.~\ref{fig:jet}.  Hence,
    in the case of the lowest-order expression for $\bar
    J^\textrm{sub}$, the IR divergences in Eq.~(\ref{eq:soft-e+f})
    cancel those in Eq.~(\ref{eq:jet-e+f}). As we have already mentioned
    in Sec.~\ref{sec:J-A-alphas}, this cancellation does not hold in
    general because Eq.~(\ref{eq:jet-e+f}) is missing some of the
    radiative-jet-function contributions whose IR divergences would be canceled by
    the soft-subtraction contributions in Eq.~(\ref{eq:soft-e+f}).

We can also see that, in the case of the lowest-order expression for
$\bar J^\textrm{sub}$, the nonlocal contributions in
Eq.~(\ref{eq:soft-e+f}) vanish. However, as we have remarked earlier,
this feature is special to the lowest-order case.

Owing to the presence of the Wilson-line denominator $k_-+i\ve$, the
expressions in Eqs.~(\ref{eq:jet-e+f-split-b}) and (\ref{eq:soft-e+f})
develop rapidity divergences as $k_+/k_-\to \infty$.  These rapidity
divergences cancel in the difference between
Eqs.~(\ref{eq:jet-e+f-split-b}) and (\ref{eq:soft-e+f}). Therefore, in
order to avoid introducing a rapidity regulator, we compute the
expressions in Eqs.~(\ref{eq:jet-e+f-split-b}) and (\ref{eq:soft-e+f})
together. We have organized the expression in the third line of
    Eq.~(\ref{eq:soft-e+f}) in such a way as to make the cancellation of
    the rapidity divergences explicit.

For the jet and soft-subtraction expressions in
Eqs.~(\ref{eq:jet-e+f-split}) and (\ref{eq:soft-e+f}), we perform the
$k_{-}$ contour integration first.  Carrying out the $k_{-}$ contour
integration by deforming the contour into the lower half-plane (making
the assumption $p_+>0$), we obtain
\begin{subequations}
\begin{eqnarray}
&&
\alpha_s
(-Z_{\bar J}^{(1)}\otimes \bar J^{\rm sub})_{(\textrm{e+f}),1}
\nonumber \\
&=&
g_{s}^{2}C_{F}\left(\frac{\mu^{2}e^{\gamma_{E}}}{4\pi}
\right)^{\epsilon}
\Bigg\{
\int \frac{d^{D-2}
k_{\perp}}{(2\pi)^{D-2}}
\int_{0}^{p_+}\frac{dk_{+}}{2\pi} \frac{\left(1+\e\frac{p_{+}}{k_{+}}\right)
\bar{J}^{\textrm{sub}}(p^{2})}
{(p^2-p_-k_+-\frac{p_+}{k_+}\bm{k}_\perp^2+i\varepsilon)
p_{+}}
\Bigg\}_\textrm{UV},
\\[1ex]
&&
\alpha_s
(-Z_{\bar J}^{(1)}\otimes \bar J^{\rm sub})_{(\textrm{e+f}),2}
\nonumber \\
&=&
g_{s}^{2}C_{F}\left(\frac{\mu^{2}e^{\gamma_{E}}}{4\pi}
\right)^{\epsilon}
\Bigg\{
\int \frac{d^{D-2} k_{\perp}}{(2\pi)^{D-2}}
\Bigg[
-\int_0^{p_+} \frac{dk_{+}}{2\pi} \frac{p_{-}k_{+}}
{(p^2-p_-k_+-\frac{p_+}{k_+}\bm{k}_\perp^2+i\varepsilon)
p_{+}\bs{k}_{\perp}^{2}}\nn\\
&&
\quad\quad\quad\quad\quad\quad\quad\quad\quad\quad\quad\quad\quad
+\int^{p_{+}}_{-\infty} \frac{dk_{+}}{2\pi}\frac{p_{-}}
{\bs{k}_{\perp}^{2}(p^2-p_-k_+-\bm{k}_\perp^2+i\varepsilon)}\Bigg]
\bar{J}^{\textrm{sub}}(p^{2})
\Bigg\}_\textrm{UV},
\\[1ex]
&&
\alpha_s
(-Z_{\bar S}^{(1)}\otimes \bar J^{\rm sub})_{(\textrm{e+f})}
\nonumber \\
&=&
g_{s}^{2}C_{F}\left(\frac{\mu^{2}e^{\gamma_{E}}}{4\pi}\right)^{\epsilon}
\Bigg\{
\int \frac{d^{D-2} k_{\perp}}{(2\pi)^{D-2}}
\int^{p_{+}}_{-\infty} \frac{dk_{+}}{2\pi}
\bigg[-\frac{k_{+}p_{-}\theta(k_{+})}
{p_{+}\bs{k}_{\perp}^{2}(p^{2}-p_{-}k_{+}+i\varepsilon)}
+\frac{p_{-}}{\bs{k}_{\perp}^{2}(p^{2}-p_{-}k_{+}+i\varepsilon)}\bigg]
\nonumber \\
&&
\quad\quad\quad\quad\quad\quad\quad
\times
\left[\bar{J}^{\textrm{sub}}(p^{2})
+ \frac{p^{2}\bar{J}^{\textrm{sub}}(p^{2})-
(p^{2}-k_{+}p_{-})\bar{J}^{\textrm{sub}}(p^{2}-k_{+}p_{-})}
{-k_{+}p_{-}+i\varepsilon}\right]
\Bigg\}_\textrm{UV}.
\end{eqnarray}
\end{subequations}
Note that the expression for 
$(-Z_{\bar S}^{(1)}\otimes \bar J^{\rm sub})_{(\textrm{e+f})}$ 
contains a scaleless
    integral over $k_\perp$ that would produce a factor $1/\epsilon_{\rm
      UV}-1/\epsilon_{\rm IR}$. That is, the soft subtraction again
    vanishes in dimensional regularization if one does not distinguish
    UV and IR poles, and its role is to convert nonlocal IR poles to
    nonlocal UV poles.

Using Eq.~(\ref{eq:alphas-renorm}), making the changes of variables
$\bs{k}_{\perp}^{2} =t p^{2}$ and $k_{+}=(1-x)p_{+}$, and using
$p_+p_-\approx p^2$, we find that the
total contribution to $(-Z_{\bar J^{\rm sub}}^{(1)}\otimes \bar J^{\rm
  sub})_{G}$ is
\begin{eqnarray}
&&
\alpha_s
(-Z_{\bar J^{\rm sub}}^{(1)}\otimes 
\bar J^{\rm sub})_{G}
\nonumber \\
&=&
\alpha_s
\left[
(-Z_{\bar J}^{(1)}\otimes \bar J^{\rm sub})_{(\textrm{e+f}),1}
+
(-Z_{\bar J}^{(1)}\otimes \bar J^{\rm sub})_{(\textrm{e+f}),2}
-
(-Z_{\bar S}^{(1)}\otimes \bar J^{\rm sub})_{(\textrm{e+f})}
\right]
\nonumber \\
&=& \frac{\alpha_{s}C_{F}}{2\pi}\left(\frac{\mu^{2}e^{\gamma_{E}}}{p^{2}}\right)^{\epsilon}\int_{0}^{\infty} 
\frac{dt\,t^{-\e}}{\Gamma(1-\e)}
\nonumber \\
&&
\times
\Bigg\{ \int_{0}^{1}dx
\frac{(-\e -1+x)\bar{J}^{\textrm{sub}}(p^{2})}{t-x(1-x)-i\varepsilon}
-\int_{0}^{\infty}d x\frac{1}{t} 
\bigg(\frac{1}{t-x-i\varepsilon}+\frac{1}{x}\bigg)
\bar{J}^{\textrm{sub}}(p^{2})
\nn\\
&&
\quad
+\int_{0}^{1}d x\frac{1}{t}
\bigg[\frac{(1-x)^{2}}{t-x(1-x)-i\varepsilon}+\frac{1-x}{x}\bigg]
\bar{J}^{\textrm{sub}}(p^{2})\nn\\
&&
\quad
-\int_{0}^{\infty}d x\frac{1}{tx}\frac{\bar{J}^{\textrm{sub}}(p^{2})-x\bar{J}^{\textrm{sub}}(xp^{2})}{x-1}
-\int_{0}^{1}d x\frac{1}{tx}\left[\bar{J}^{\textrm{sub}}(p^{2})-x\bar{J}^{\textrm{sub}}(xp^{2})\right]
\Bigg\}_\textrm{UV}.
\end{eqnarray}
Completing the $t$ integration, we obtain
\begin{eqnarray}
\label{eq:JGsub}
&&
\alpha_s
(-Z_{\bar J^{\rm sub}}^{(1)}\otimes 
\bar J^{\rm sub})_{G}
\nonumber \\
&=& \frac{\alpha_{s}C_{F}}{2\pi}\Bigg\{\left(\frac{\mu^{2}e^{\gamma_{E}}}{-p^{2}-i\varepsilon}\right)^{\epsilon}\Gamma(\e_{\textrm{UV}})\bigg[\int_{0}^{1}d x\frac{x-1-\e}{\left[x\left(1-x\right)\right]^{\e}}+\int_{0}^{1}d x \frac{(1-x)^{1-\e}-1}{x^{1+\e}}-\int_{1}^{\infty} \frac{d x}{x^{1+\e}}\bigg]\bar{J}^{\textrm{sub}}(p^{2}) 
\nonumber \\
&&
\quad\quad
+ \left(\frac{1}{\e_{\textrm{UV}}}-\frac{1}{\e_{\textrm{IR}}}\right)
\left[
\int_{1}^{\infty}dx
\frac{
x\bar{J}^{\textrm{sub}}(xp^{2})
-\bar{J}^{\textrm{sub}}(p^{2})
}
{x(x-1)}
-
\int_{0}^{1}dx
\frac{
x\bar{J}^{\textrm{sub}}(xp^{2})
-\bar{J}^{\textrm{sub}}(p^{2})
}{1-x}
\right]
\Bigg\}_\textrm{UV}
\nn\\
&=& 
\frac{\alpha_{s}C_{F}}{4\pi}\Bigg\{\left(\frac{\mu^2e^{\gamma_{E}}}{-p^{2}-i\varepsilon}\right)^{\epsilon}\frac{\Gamma(1+\e)\Gamma^2(-\e_{\textrm{UV}})}{\Gamma(2-2\e)}\left(-2+\e -2\e^{2}\right)\bar{J}^{\textrm{sub}}(p^{2})
\nn\\
&&
\quad\quad
+ 2\left(\frac{1}{\e_{\textrm{UV}}}-\frac{1}{\e_{\textrm{IR}}}\right)\left[\int_{1}^{\infty}dx\frac{x\bar{J}^{\textrm{sub}}(xp^{2})}
{[x(x-1)]_{+}}
-\int_{0}^{1}dx\frac{x\bar{J}^{\textrm{sub}}(xp^{2})}{[1-x]_{+}}\right]\Bigg\}
_\textrm{UV},
\end{eqnarray}
where we have arranged this expression so as to make the cancellation of
    the rapidity divergences explicit. Finally, extracting the UV
divergences, we have
\begin{eqnarray}
\alpha_s
(-Z_{\bar J^{\rm sub}}^{(1)}\otimes \bar J^{\rm sub})_{G}
&=&  \frac{\alpha_{s}C_{F}}{4\pi}\Bigg\{\left[-\frac{2}{\e_{\textrm{UV}}^{2}}-\frac{2}{\e_{\textrm{UV}}}\ln\left(\frac{\mu^{2}}{-p^{2}-i\varepsilon}\right)-\frac{3}{\e_{\textrm{UV}}}\right]\bar{J}^{\textrm{sub}}(p^{2})\nn\\
&&
\quad\quad\quad
+\frac{2}{\e_{\textrm{UV}}}\left[\int_{1}^{\infty}dx\frac{x\bar{J}^{\textrm{sub}}(xp^{2})}{[x(x-1)]_{+}}
-\int_{0}^{1}dx\frac{x\bar{J}^{\textrm{sub}}(xp^{2})}{[1-x]_{+}}\right]\Bigg\}.
\end{eqnarray}

%%%%%%%%%%%%%%%%%%%%%%%%%%%%%%%%%%%%%%%%%%%%%%
%%%%%%%%%%%%%%%%%%%%%%%%%%%%%%%%%%%%%%%%%%%%%%%
\subsection{Renormalization of the subtracted jet function $\pmb{\bar{J}}^{\textbf{sub}}$(\textbf{\textit{p}}$^{\textbf{2}}$) at order $\pmb{\alpha_s}$ }
The total of the UV divergences at order $\alpha_{s}$ is
\begin{eqnarray}
\alpha_s
(-Z_{\bar J^{\rm sub}}^{(1)}\otimes \bar J^{\rm sub})
&=&
\alpha_s
(-Z_{\bar J^{\rm sub}}^{(1)}\otimes \bar J^{\rm sub})_{A}
+
\alpha_s
(-Z_{\bar J^{\rm sub}}^{(1)}\otimes \bar J^{\rm sub})_{G}
\nn\\
&=&\frac{\alpha_{s}C_{F}}{4\pi}\Bigg\{\left[\frac{2}{\e_{\textrm{UV}}^{2}}+\frac{2}{\e_{\textrm{UV}}}\ln\left(\frac{\mu^{2}}{-p^{2}-i\varepsilon}\right)\right]\bar{J}^{\textrm{sub}}(p^{2})\nn\\
&&
\quad\quad\quad
-\frac{2}{\e_{\textrm{UV}}}\left[\int_{1}^{\infty}dx\frac{x\bar{J}^{\textrm{sub}}(xp^{2})}{[x(x-1)]_{+}}+\int_{0}^{1}dx\frac{x\bar{J}^{\textrm{sub}}(xp^{2})}{[1-x]_{+}}\right]\Bigg\}.
\end{eqnarray}

From Eq.~(\ref{eq:Z-J-sub}), it then follows that
    $Z_{J}^{\rm sub}[(1-x),p^{2};\mu]$, up to order $\alpha_{s}$, is given by
\begin{subequations}%
\label{eq:as-anom-dim}%
\begin{eqnarray}
Z_{J}^{\rm sub}[(1-x),p^2;\mu]&=&
  \left[1+\frac{\alpha_{s}C_{F}}{4\pi}
\left(-\frac{2}{\e^{2}}-\frac{2}{\e}
\ln\frac{\mu^{2}}{-p^{2}-i\varepsilon}\right)\right]\delta(1-x)\nn\\ 
&&\qquad + \frac{\alpha_{s}C_{F}}{2\pi\e}\Gamma(1,x)\theta(x),
\phantom{X}
\end{eqnarray}
where
\begin{eqnarray}
\Gamma(1,x)=\left[\frac{\theta(1-x)}{1-x}+\frac{\theta(x-1)}{x(x-1)}\right]_{+}.
\end{eqnarray}
\end{subequations}%
This result is in agreement with Eq.~(2.6) of Ref.~\cite{Liu:2020ydl}
for the case $y=1$.\footnote{Note that, $Z_J^{\rm
          sub}[(1-x),p^2;\mu]=Z_J^{\rm LN}(p^2,xp^2,;\mu)$, where $Z_J^{\rm
          LN}(p^2,xp^2;\mu)$ is the quantity that appears in
        Ref.~\cite{Liu:2020ydl}.}  Therefore, our result for the
renormalization kernel of the radiative jet function with soft
subtractions agrees with the one-loop result that had been inferred from
the factorization theorem for $B\to \gamma \ell^-\nu$
\cite{Bosch:2003fc}.  Note, however, that, in Ref.~\cite{Bosch:2003fc},
the renormalization kernel was ascribed to $\bar J$, rather than to
$\bar J^{\rm sub}$.

The result in Eq.~(\ref{eq:as-anom-dim}) is compatible with the
    analyticity properties of $\bar J$, which is analytic in the $p^2$
    upper half-plane \cite{Liu:2019oav}. Hence, although we have derived this
    result for the time-like case $p^2>0$, it can be continued
    analytically to the space-like case $p^2<0$. We have also checked
    the result for the space-like case by explicit calculation.

\section{Vanishing of the soft subtractions in dimensional
  regularization}
  \label{sec:softvanish}

As we have noted in Secs.~\ref{sec:J-A-alphas} and
    \ref{sec:J-G-alphas}, the order-$\alpha_s$ soft subtractions vanish
    in dimensional regularization if one does not distinguish between UV
    and IR poles. We now argue that this is a general property of the
    soft subtractions at all orders in $\alpha_s$.

In the soft subtractions, the only propagator denominators that contain
a scale are those that are associated with the $n$-hard-collinear-soft
quark line. In these denominators, the scale-dependent terms have the
form $k_{i+} p_-$, where $k_i$ is a loop momentum. If we rescale all of
the loop momenta according to
\begin{eqnarray}
k_{i-}&\to& k_{i-}\rho^2,
\nn\\ 
k_{i\perp}&\to& k_{i\perp}\rho,
\nn\\
k_{i+}&\to& k_{i+}\rho^0,
\end{eqnarray}
then the resulting expression is homogeneous in $\rho$. Therefore, the
soft subtractions are scaleless integrals that vanish in dimensional
regularization if one does not distinguish between UV and IR
divergences. Hence, the soft subtractions do not affect the results of
the existing fixed-order calculations of the radiative jet function
\cite{Liu:2019oav,Liu:2020ydl}.
  
\section{Summary and discussion}
\label{sec:summary}

The radiative jet function is a quantity that appears in the
factorization theorems for the exclusive processes $B\to \gamma\ell\nu$
\cite{Bosch:2003fc} and $H\to\gamma\gamma$ through a $b$-quark loop
\cite{Liu:2019oav}. Its renormalization-group evolution is an
essential ingredient in the resummation of logarithms of $m_b/\mu$ and
$m_H/m_b$ in these processes.
  
Notwithstanding the importance of these applications, no
    direct calculation of the renormalization-group kernel for
    the radiative jet function exists in the literature. Rather, the
    renormalization-group kernel has been inferred from the
    factorization theorem for $B\to\gamma\ell\nu$ and the known
    renormalization-group kernel for the $B$-meson light-front
    distribution \cite{Bosch:2003fc}.

In this paper, we have argued that the radiative jet function contains,
in addition to hard-collinear contributions, soft contributions that
must be subtracted in order to avoid double counting of soft
contributions that appear in other quantities in the factorization
theorems. These soft subtractions are zero-bin subtractions in the
language of SCET.  We have shown that the radiative jet function can be
factored into a convolution over a light-front momentum of a
soft-subtraction function and a subtracted radiative jet function. The
subtracted radiative jet function is free of soft divergences, which are
contained entirely in the soft-subtraction function.  It is the
subtracted jet function that should properly appear in the factorization
theorems.

The renormalization-group kernel of the subtracted radiative jet
function derives from two sources: (1) the renormalization-group kernel
of the radiative jet function, which is local and (2) the
renormalization-group kernel of the soft-subtraction function, which
is nonlocal and leads to the nonlocal contributions in the
    renormalization-group kernel of the subtracted radiative jet
    function.

We have shown that the soft-subtraction contributions contain scaleless
integrals in dimensional regularization. That is, they are proportional
to $1/\epsilon_{\textrm{UV}}-1/\epsilon_{\textrm{IR}}$, and they vanish
in calculations in which one does not distinguish UV and IR
poles. Hence, existing fixed-order calculations of the radiative jet
function are not changed by the soft subtractions, which merely convert
IR poles to UV poles. This is the usual role of soft (zero-bin)
subtractions in factorization theorems. However, because the UV (and IR)
divergences that arise from the soft subtractions are nonlocal beyond
one-loop order, the soft subtractions give a nonlocal contribution
to the renormalization-group kernel of the subtracted radiative jet
function. To the best of our knowledge, this is a novel phenomenon.

We have illustrated the role of soft subtractions in the renormalization
of the subtracted radiative jet function by carrying out a complete
calculation of the renormalization-group kernel in order $\alpha_s$. Our
result agrees with the renormalization-group kernel that was inferred in
Ref.~\cite{Bosch:2003fc}. However, in that work, the
    renormalization-group kernel was ascribed to the radiative jet
    function, rather than to the subtracted radiative jet function.
The renormalization-group kernel in order $\alpha_s^2$ has also been
inferred from the factorization theorem for $B\to\gamma\ell\nu$
    \cite{Liu:2020ydl}. It would be interesting to verify that
analysis by making use of the methods that we have presented for the
direct calculation of the renormalization-group kernel.  However, that
work is beyond the scope of the present paper.

There is a large class of exclusive processes that proceed at
    subleading power in the hard-scattering scale \cite{Moult:2019mog}.
    Helicity-flip processes whose amplitudes contain singularities at
    the endpoints of light-cone distribution amplitudes are included in
    this class.  While factorization theorems have not yet been worked
    out in detail for most subleading-power processes, it is known that
    jet functions are a general feature of the factorization theorems
    \cite{Moult:2019mog}. Since collinear functions, including jet
    functions at subleading power, generally contain soft contributions
    that must be subtracted in order to avoid double counting, we expect
    that our methods would be useful in working out the
    renormalization/evolution of such jet functions.

In particular, there is a jet function that is identical to the one in
    Eq.~(\ref{defJF}), except that the external photon state is replaced
    by an external gluon state. This jet function would be relevant, for
    example, in the factorization theorem for $H\to gg$ through a
    $b$-quark loop. We would expect the analysis of soft subtractions in
    the present paper to go through essentially unchanged in this case
    because soft-gluon attachments to the external gluon would
    compensate in the graphical Ward identities for the fact that the
    external gluon, unlike the external photon, carries color.

The soft function that appears in the factorization theorem for
$H\to\gamma\gamma$ \cite{Liu:2019oav,Bodwin:2021cpx} also has a nonlocal
renormalization kernel. It has been conjectured in
Ref.~\cite{Liu:2020eqe} that the complete soft sector in the
factorization theorem, which includes contributions from the soft-quark
function, two radiative jet functions, and a short-distance Wilson
coefficient, is a renormalization-group invariant. That conjecture leads
to a prediction for the renormalization-group kernel of the soft
function \cite{Bauer:2000yr}, and that prediction was verified in order
$\alpha_s$ in Ref.~\cite{Bodwin:2021cpx}.  While the soft-subtraction
function that we have defined in the present paper clearly has
similarities with the soft function in the factorization theorem, it is
not obvious that they would combine to cancel the nonlocal contributions
to the renormalization-group kernel for the complete soft sector. It
would be interesting to explore systematically the issue of the
renormalization-group invariance of the soft sector by making use of the
expression for the subtracted radiative jet function that we have
presented in this paper.

\begin{acknowledgments}

We wish to acknowledge the contributions of Hee Sok Chung to work on
related topics at an early stage in this project.  We thank Ze Long Liu
and Jian Wang for several useful discussions.  The work of G.T.B.\ and
X.-P.W.\ is supported by the U.S.\ Department of Energy, Division of
High Energy Physics, under Contract No.\ DE-AC02-06CH11357.  The work of
J.L. is supported by the National Research Foundation of Korea under
Contract No. NRF-2020R1A2C3009918.  The work of J.-H.E. is supported by
the National Natural Science Foundation of China (NSFC) through Grant
No. 11875112.  The submitted manuscript has been created in part by
UChicago Argonne, LLC, Operator of Argonne National Laboratory. Argonne,
a U.S.\ Department of Energy Office of Science laboratory, is operated
under Contract No.\ DE-AC02y-06CH11357. The U.S.\ Government retains for
itself, and others acting on its behalf, a paid-up nonexclusive,
irrevocable worldwide license in said article to reproduce, prepare
derivative works, distribute copies to the public, and perform publicly
and display publicly, by or on behalf of the Government.

All authors contributed equally to this work.

\end{acknowledgments}

\appendix

\section{Re-arrangement of the blob in the second diagram of
  Fig.~\ref{fig:jetsub}\label{app:blob-rearrangement}}

In this appendix, we sketch, in the diagrammatic approach, the
re-arrangement of the blob in the second diagram of
Fig.~\ref{fig:jetsub} into the form of a radiative jet function (with
    soft subtractions). The issue here is the disposition of gluons
    that attach to the quark line to the left of the real-photon
    vertex.

First, we note, as we have mentioned previously, that we have
defined $\bar J$ in Eq.~(\ref{defJF}) in such a way that the vertex for
a transverse photon is always $ie_{q}\gamma_\perp^\mu$, regardless of
whether the photon attaches to a quark line or to a covariant
derivative. Hence, we can treat all real-photon vertices uniformly,
    regardless of their origin.

Next, we consider an $n$-hard-collinear gluon that connects an
$n$-hard-collinear line to the quark line at a point to the left of the
real-photon vertex. We write the numerator factor of this gluon as
$\gamma\cdot j$, where the Dirac matrix $\gamma$ is on the quark
line, and $j$ is the current to which the other end of the gluon
attaches. Note that $j$ is the standard QCD current minus the
Grammer-Yennie form in the case that there is a soft subtraction
associated with the gluon.

We decompose $\gamma\cdot j$ into its light-cone components:
\begin{equation}
\label{eq:vertex-decomp}
\gamma\cdot j= \frac{1}{2}\gamma\cdot n \bar n \cdot j 
+\frac{1}{2}\gamma\cdot \bar n n\cdot j +\gamma_\perp \cdot j_\perp.
\end{equation}
We can further decompose the first term in Eq.~(\ref{eq:vertex-decomp})
by making use of the identity
\begin{eqnarray}
\label{eq:decomp2}
\frac{1}{2}\gamma\cdot n \bar n\cdot j&=&\left(\gamma\cdot k
 -\gamma_{\perp}\cdot k_\perp
-\frac{1}{2}\gamma\cdot \bar n n\cdot k\right )
\frac{\bar n\cdot j}{\bar n\cdot k},
\end{eqnarray}
where $k$ is the momentum of the gluon. Re-arranging terms, we obtain
\begin{eqnarray}
\label{eq:re-arrange}
\gamma\cdot j= \gamma\cdot k \frac{\bar n \cdot j}{\bar
  n\cdot k} 
+\left(\gamma_\perp\cdot j_\perp-
\gamma_\perp\cdot k_\perp\frac{\bar n\cdot j}{\bar n\cdot k}\right)
+\frac{1}{2} \gamma\cdot \bar n n\cdot j
-\frac{1}{2} \gamma\cdot \bar n n\cdot k 
\frac{\bar n\cdot j}{\bar n \cdot k}.
\end{eqnarray}
The first term in Eq.~(\ref{eq:re-arrange}) gives the longitudinally
polarized (pure gauge) part of the gluon numerator. The terms in
parentheses give a transverse-covariant-derivative contribution,
where the first term in parentheses is the gauge-field part of the
covariant derivative and the second term in parentheses is the
ordinary-derivative part of the covariant derivative times a Wilson-line
contribution. The last two terms give contributions that are suppressed
as $\lambda$ if the left-hand connection of the gluon to the quark line
is at the transition on the quark line between a soft momentum and
an $n$-hard-collinear momentum.

Suppose that some number of longitudinally polarized gluons attach to
    the quark line to the left of the real-photon attachment. These can
    be re-arranged by making use of the graphical Ward identities into
    attachments to a Wilson line $W_n^\dagger$.  If there are no
    transverse-covariant-derivative connections to the quark line to the
    left of the real-photon attachment, then the Wilson line
    $W_n^\dagger$ attaches to the quark line just to the left of the
    real-photon attachment. Otherwise, the Wilson line $W_n^\dagger$
    attaches to the quark line just to the left of the left-most
    transverse-covariant-derivative connection. As we have mentioned, at
    this connection to the quark line, the contributions of the last two
    terms in Eq.~(\ref{eq:re-arrange}) are suppressed as
    $\lambda$. However, for connections to the quark line to the right
    of this connection, the last two terms in Eq.~(\ref{eq:re-arrange})
    can contribute in leading power in $\lambda$, and so the complete
    QCD expression for $\gamma\cdot j$ must be kept for gluons with
    these connections to the quark line.

Finally, we insert a projector $P_n$ to the left of the entire
expression to project out the components of the quark field that are
large under $n$-hard-collinear scaling. The motivation for this is that
we wish to retain only the component of the quark field that
corresponds to the effective field theory for the $n$-hard-collinear
sector. The presence of $P_n$ insures that the blob cannot contain
wrong-collinearity contributions.

At this point, the blob in the second diagram of Fig.~\ref{fig:jetsub}
is in the form of a radiative jet function (with soft subtractions).

%\nocite{*}

%\bibliography{covariant}% Produces the bibliography via BibTeX.

\end{document}